\documentclass{aa}  

\usepackage{graphicx}
\usepackage{txfonts}
\usepackage{lipsum}
\usepackage{lscape}
\usepackage{placeins}
\usepackage{natbib}
\bibpunct{(}{)}{;}{a}{}{,}                      
\usepackage{hyperref}

\begin{document}

   \title{CoBiToM project}
   \subtitle{III. Physical characterisation of the potential merger candidates AL~Lep, ASAS~J082243+1927.0, and ZZ~PsA}

   \titlerunning{CoBiToM project - III. Potential merger candidates.} 
   \authorrunning{Palafouta et al.}

   \author{S. Palafouta\inst{1}\corrauth{spalafou@phys.uoa.gr}
        \and K. D. Gazeas\inst{1}\email{kgaze@phys.uoa.gr}
        \and P. Hakala\inst{2}\email{pahakala@astro.helsinki.fi}
        \and S. Zola\inst{3}\email{szola@oa.uj.edu.pl}
        \and A. Kafkas\inst{1}\email{sph2100239@uoa.gr}
        \and M. Garro\inst{1}\email{mariagarro321@gmail.com}
        }

   \institute{Section of Astrophysics, Astronomy and Mechanics, Department of Physics, National and Kapodistrian University of Athens, GR-15784 Zografos, Athens, Greece
   \and Finnish Centre for Astronomy with ESO (FINCA), University of Turku, FI-20014, Turku, Finland
   \and Astronomical Observatory of the Jagiellonian University, ul. Orla 171, 30-244 Krakow, Poland}

   \date{Received June 12, 2026}
 
  \abstract
   {Contact binaries (CBs) are key systems for investigating the late evolutionary stages that may lead to stellar coalescence. Systems with an extremely low mass ratio and high fill-out factor are of particular interest as potential merger candidates and possible progenitors of luminous red novae.}
   {A combined photometric and spectroscopic study of the CB systems AL~Lep, ASAS~J082243+1927.0, and ZZ~PsA is presented. The aim is to derive self-consistent orbital and physical parameters, assess their evolutionary state, and examine whether they have reached the terminal and unstable phase of their evolution.} 
   {High-resolution VLT/UVES spectra were analysed to derive radial velocity values, spectroscopic mass ratio, and metallicity. The results were combined with multi-band light-curve modelling within a Monte Carlo (MC) framework to determine the system geometries and absolute parameters. Orbital period variations were investigated through $O-C$ analysis, while the evolutionary status of the systems was assessed using the Darwin instability criterion, critical mass ratio estimates, fill-out factor, and orbital period variation. The derived parameters were also compared with PARSEC zero-age main sequence (ZAMS) and terminal-age main sequence (TAMS) reference curves, selected according to the metallicity of each system.}
   {Mass ratio values of $q=0.1332$, $0.1041$, and $0.0789$ were derived for AL~Lep, ASAS~J082243+1927.0, and ZZ~PsA, respectively. The corresponding primary mass values are $M_1=1.656\pm0.081~M_{\odot}$, $1.275\pm0.078~M_{\odot}$, and $1.795\pm0.111~M_{\odot}$. The locations of their components show significant deviations from single-star ZAMS and TAMS reference curves, indicating strong binary interaction and energy redistribution through the common envelope. The spin-to-orbital angular momentum ratio for AL~Lep and ASAS~J082243+1927.0 remains below the standard Darwin instability threshold, while ZZ~PsA lies close to the instability regime and may be consistent with an unstable configuration within the derived uncertainties.}
   {}


   \maketitle
   \nolinenumbers

\section{Introduction}
Luminous red novae (LRNe) are rare optical transients that are generally associated with the coalescence of low-mass binary systems. Among them, V1309~Sco remains the only event with detailed pre-outburst observations, providing a direct observational link between W~UMa type contact binaries (CBs) and stellar mergers \citep{nakano2008,tylenda2011}. Archival photometry showed that, before eruption, V1309~Sco was a rapidly evolving CB with a decreasing orbital period, offering the first direct evidence that CB evolution can culminate in a merger event \citep{tylenda2011,tylenda2013}.

Low-mass stars, typically defined as main sequence (MS) stars with a mass of below one $M_{\odot}$, comprise the majority of the Galactic stellar population, accounting for approximately $70\%$ of stars in the Milky Way \citep{kroupa2001,chabrier2003,morales2007}. Contact binaries are an important subset of this population, with statistical estimates suggesting that roughly one in 500 FGK type stars in the Galactic disk belongs to a CB system \citep{rucinski2002a,rucinski2006}. Therefore, low-mass stellar mergers represent an important channel of binary evolution and are widely considered potential progenitors of LRNe \citep{tylenda2011,pejcha2014}. Despite the observational prevalence of CBs, the physical conditions that lead a system to coalescence and produce an LRN remain incompletely understood. Low-mass stellar mergers are also utilised for explaining several stellar populations, as they provide a natural formation channel. Binary merger scenarios have been proposed as one of the main formation pathways for blue stragglers (BSSs) \citep{chen2008,sills2009,perets2009}, yellow stragglers (YSSs) \citep{leiner2016,mathieu2025}, and fast-rotating stars \citep{gaz2008,step2012,jiang2013,shepard2022} such as blue lurkers (BLs) \citep{jadhav2026}. 

Contact binaries with a very low mass ratio (LMR) and high fill-out factor, typically \(f \sim 50\%\), are considered among the most promising progenitors of red nova events \citep{rasio1995a, li2006,loukaidou2022}. Their evolution depends on mass transfer and angular momentum loss (AML), which can lead the systems towards Darwin instability when the spin angular momentum (AM) exceeds a critical fraction of the orbital AM \citep{Darwin1879,hut1980,rasio1995,rasio1995a,egg2001,step2012,nandez2014}. Such systems are therefore important for understanding both LRN progenitors and BSS formation in different stellar environments. Several studies have estimated the critical instability mass ratio for CBs under different assumptions \citep[e.g.][]{li2006, arbutina2007, arbutina2009, jiang2010, yang2015}. This condition is more easily satisfied in systems with an extremely low mass ratio \citep{wadhwa2021, wadhwa2024}, making LMR CBs prime candidates for an LRN. In addition to tidal instability, thermal instability has also been proposed as a possible merging triggering mechanism. In this scenario, the primary evolves away from the MS after central hydrogen exhaustion, while the secondary is unable to maintain a stable thermal-equilibrium configuration \citep{webbink1976}. Recent work by \citet{jiang2026} suggests that tidal instability may dominate for most W~UMa type CBs, while thermal instability may account for a smaller fraction of systems. Nevertheless, the precise mechanism that drives a CB from stable mass transfer to catastrophic coalescence remains an open question.

\begin{table*}[!t]
\caption{Basic properties of the target systems ASAS~J082243+1927.0, AL~Lep, and ZZ~PsA.}
\centering
\begin{tabular}{lccccc}
\hline\hline
System & RA & Dec & $V$ & $T_{0}$ & Period \\
       & (J2000) & (J2000) & (mag) & (HJD) & (d) \\
\hline
AL~Lep              & 05 06 17.658 & $-$20 07 52.79 & 9.67  & 2459565.39160(83) & 0.44864570(27) \\
ASAS~J082243+1927.0 & 08 22 42.998 & $+$19 26 58.40 & 11.49 & 2459638.31759(5)  & 0.28003792(1)  \\
ZZ~PsA              & 21 50 35.175 & $-$27 48 35.48 & 9.32  & 2459526.26743(30) & 0.37389506(6)  \\
\hline
\end{tabular}
\tablefoot{The $V$ magnitude is derived from APASS photometry \citep{henden2016}.}
\label{table_gen}
\end{table*}

Identifying and characterising this stellar population is essential for improving our understanding of CB evolution, stellar mergers, and red nova progenitors. This requires accurate and self-consistent determinations of the orbital and physical parameters of candidate systems. In this context, AL~Lep, ASAS~J082243+1927.0 (hereafter ASAS~J082243), and ZZ~PsA have been identified as potential stellar merger candidates within the CoBiToM project \citep{gazeas2021}. In this study, a combined analysis of photometric and spectroscopic observations is performed to constrain the geometry, mass ratio, temperature, and absolute stellar parameters of the systems \citep{gazeas2021b} and to assess their evolutionary status and possible proximity to coalescence.

\section{Target selection and data acquisition}
\subsection{Target selection}

AL~Lep was first reported as a variable star by \citet{strohmeier1967} and was much later included in the All Sky Automated Survey (ASAS; \citealt{pojmanski2002}), which provided the first high-quality observations of the system. According to SIMBAD database, AL~Lep belongs to a G0 spectral class, indicating a temperature of 5916~K, while the Tycho photometric colour index $B-V$ indicates a F7 spectral type, according to \citet{popper1980}. \citet{wadhwa2005} performed an analysis of V~band light curves (LCs) of AL~Lep using the Wilson-Devinney (WD) code and determined the physical parameters of the system. With the q-search method of \citet{maceroni1983}, he determined the mass ratio value as q=0.120(4). Later, \citet{yang2015} collected a sample of over-contact CBs, including AL~Lep, and performed a statistical analysis to find relations between their physical parameters. The system is included in the statistical study of individually studied W~UMa stars by \citet{latkovic2021}, as well as in the LMR sample of \citet{christopoulou2022}. It is also listed in the Transiting Exoplanet Survey Satellite (TESS) eclipsing binary (EB) catalogue of \citet{prsa2022} and in the Gaia DR3 EB catalogue of \citet{mowlavi2023}. More recently, AL~Lep was identified as a TESS CB candidate by \citet{ding2024}, based on observations from Sectors~5 and~32, with an orbital period of 0.448419751~d. In the following years, TESS-based analyses provided photometric solutions for the system, confirming its LMR. \citet{ding2023} derived a $q=0.153$ and $f=0.331$, while \citet{ding2025} obtained $q=0.135$ and $f=0.330$ without third light, and $q=0.120$ and $f=0.627$ when third light was included. In addition, \citet{guo2025} included AL~Lep in their sample of 84 totally eclipsing CBs observed by TESS, reporting $P=0.448645$~d and a mean temperature of about 6400~K. These previous studies indicate that AL~Lep is a LMR system, and thus a potential merger candidate. However, the system still lacks a detailed orbital characterisation and has not been comprehensively studied during the past two decades, making it a suitable target for the present study.

ASAS~J082243+1927.0 is a short-period CB system, classified as an EW-type EB. It was first discovered as a variable star by ASAS \citep{pojmanski2002} and was later included in several large photometric catalogues of contact or EBs (ROTSE-I; \citealt{gettel2006}, KELT; \citealt{pepper2008}, $BVR_{\rm C}I_{\rm C}$; \citealt{terrell2012}, ATLAS; \citealt{heinze2018}, Gaia DR3; \citealt{mowlavi2023}, TESS; \citealt{ding2024}). The system is also included in spectroscopic databases and analyses based on SDSS/APOGEE and LAMOST data \citep{jonsson2020, abdurrouf2022, sprague2022, shi2025}. The All-Sky Automated Survey for Supernovae (ASAS-SN; \citealt{shappee2014, kochanek2017, jayasinghe2018}) reported an orbital period of 0.28004~d, a mean $V$-band magnitude of 11.40~mag, and a variability amplitude of 0.27~mag. The first detailed photometric solution was presented by \citet{kandulapati2015}, who conducted a high-precision CCD photometric study and determined a mass ratio of $q\simeq0.106$ and a fill-out factor of $f \simeq 72\%$, indicating a deep contact configuration. They also reported phase-dependent variations in the H$\alpha$ line equivalent width, suggesting chromospheric activity, and proposed that the system may be close to merging, potentially forming a rapidly rotating single star. \citet{rukmini2016} presented photometric and H$\alpha$ studies of two short-period CBs with an extreme mass ratio, including ASAS~J082243+1927.0. They confirmed its LMR and fill-out factor values ($q\simeq 0.121$, $f\simeq 72\%$), supporting that ASAS~J082243+1927.0 is a deep CB with potential chromospheric activity. The system was also included in statistical samples of deep, LMR over-CBs \citep{yang2015, kjurkchieva2017, kjurkchieva2018, christopoulou2022, li2023, zhang2025}. More recently, \citet{li2024} performed a photometric, spectroscopic, and orbital period analysis of ASAS~J082243+1927.0. They derived a mass ratio of $q \simeq 0.094$ through the $q$-search method and a fill-out factor of $f \simeq 69.4\%$, confirming the system as a LMR, deep-CB. They also detected two flare events in TESS LCs, attributing them to the primary component. They reported excess H$\alpha$ emission in LAMOST spectra and derived an orbital period change rate of $\mathrm{d}P/\mathrm{d}t = -1.43(\pm 0.19) \times 10^{-7}~\mathrm{d~yr^{-1}}$ from the $O-C$ diagram. ASAS~J082243+1927.0 was further analysed by \citet{ding2025}, who obtained photometric solutions accounting for third light contribution. It was also included in the merger-timescale study of \citet{jiang2026}, who concluded that the system is expected to have experienced thermal instability. These previous studies consistently identify ASAS~J082243+1927.0 as a LMR, deep-contact system with magnetic activity and possible merger relevance. However, the available spectroscopic information is limited, while the existing photometric analyses are either based on survey data or on datasets with restricted temporal and/or filter coverage. Therefore, a combined analysis based on new photometric and spectroscopic data is required to further investigate the merger potential of the system.

ZZ~PsA was identified as a variable star by \citet{strohmeier1967} and later revisited by \citet{demartino1996}. It is included in several photometric, X-ray, and EB catalogues and surveys \citep{pojmanski2002, szczygiel2008, kiraga2012, haakonsen2009}, as well as the statistical catalogue of individually studied W~UMa systems by \citet{latkovic2021}, the TESS EB catalogue of \citet{prsa2022}, and the recent TESS study of totally eclipsing CBs by \citet{guo2025}. In the latter work, ZZ~PsA was analysed using TESS photometry and was confirmed as a LMR deep-contact system with parameters consistent with previous studies ($q \simeq 0.090$, $f \simeq 49\%$). ASAS observations confirmed the contact nature of the system with an orbital period of $0.3738933$~d, while a single time of minimum was reported by \citet{otero2004}. Preliminary photometric analysis of the ASAS data by \citet{wadhwa2006} suggested that ZZ~PsA is an extremely LMR system ($q \simeq 0.08$) with a high degree of contact ($f \sim 90\%$). The first dedicated multi-band photometric analysis of the system was presented by \citet{wadhwa2021}, who derived a mass ratio of $q = 0.078$, a fill-out factor exceeding $95\%$, and a primary-component mass of $M_1 \simeq 1.213\,M_{\odot}$, consistent with the reported F6 spectral type. In the same study, the authors revisited the instability mass ratio criterion for CBs and developed a relation that links the instability mass ratio to the primary mass and the degree of contact. Using this approach, they identified ZZ~PsA as a potential merger candidate and proposed it as a possible red-nova progenitor. The system was later included in statistical and evolutionary studies of LMR CBs and merger candidates \citep{christopoulou2022}. These previous studies identify ZZ~PsA as an extreme LMR, deep-contact system and a possible merger. However, the system still lacks detailed spectroscopic characterisation and long-term combined photometric and orbital investigation, making it a suitable target for the present analysis.


\subsection{Observatories and data acquisition}
\subsubsection{Spectroscopic data}
AL~Lep, ASAS~J082243+1927.0, and ZZ~PsA were observed with the Very Large Telescope (VLT) at Cerro Paranal, Chile, using the Ultraviolet and Visual Echelle Spectrograph (UVES) on VLT/UT2 (Kueyen). The observations yielded high-resolution spectra with a resolving power of $R\simeq 40000$. The data were acquired during multiple observing runs between November 2023 and January 2024. Spectra were obtained with the lower (blue, L) and upper (red, U) arms of UVES, covering the wavelength ranges 479-575~nm and 583-679~nm, respectively. A summary of the observational details is provided in Table~\ref{vlt_appl} in the appendix.


\subsubsection{Photometric data}
Photometric observations of all systems in this study were obtained from both ground and space observatories. Ground-based data were mainly obtained at the University of Athens Observatory (UOAO\footnote{\url{http://observatory.phys.uoa.gr}}) between 2021 and 2024 in the framework of CoBiToM Project. The facility utilises a 0.4~m (f/8) Cassegrain telescope equipped with a set of $UBVRI$ filters (Bessell specifications), and a SBIG STF-3200W CCD camera with an f/6.3 focal reducer \citep{gaz2016}. The observing setup provides a field of view of 17$\times$26 arcmin. The same setup was also used to obtain clear aperture LCs, in order to extract times of minimum light (ToM) during eclipses, and produce LCs with improved temporal accuracy. Additional ground-based observations of ASAS~J082243+1927.0 were obtained in 2024 through the Europlanet 2024 Research Infrastructure (RI) NA2 programme, utilising the reflecting telescope at Kryoneri Astronomical Station, operated by the Institute for Astronomy, Astrophysics, Space Applications and Remote Sensing of the National Observatory of Athens. This facility consists of a 1.2~m (f/2.8) prime-focus Cassegrain telescope, equipped with $R$ and $I$ filters (Johnson-Cousins specifications), and two sCMOS Andor Zyla 5.5 cameras \citep{xilouris2018}. These observations were used to derive ToM light, which are required for the O-C analysis and the study of orbital period variation. The ground-based observing strategy was designed to obtain multicolour photometric data for all systems using the same instrumental setup for each target. This approach minimises possible systematic effects due to instrumental calibration differences or filter mismatches. In addition, complete LCs were obtained over the shortest possible time interval, in order to reduce the impact of intrinsic variability, such as magnetic activity, which could prevent the combination of data from different observing runs. 

TESS provides nearly continuous space-based photometry, making it suitable for constructing broad-band photometric models. Therefore, TESS Science Processing Operations Center (SPOC) data products \citep{caldwell2020} were used for both the LC modelling and ToM extraction. The adopted datasets were downloaded using the Python package 'lightkurve' \citep{lightkurve2018} and the relevant details are presented in Table~\ref{tab:phot_obs_summary}. Since TESS observes in a broad, non-standard passband, these LCs were analysed separately from the ground-based multi-band photometry. Additional ToM were derived from the SuperWASP (SWASP; \citealt{poll2006}) database, and collected from online minima databases\footnote{\url{http://var.astro.cz/en/}}. Other available photometric datasets were also examined, but were not included in the LC analysis due to their sparse sampling, making them less suitable for accurate time-domain analysis.

The observational time span and corresponding observing details for each of the three systems are summarised in Table~\ref{tab:phot_obs_summary}. For AL~Lep, the B, V, R, I, and C observations, obtained between August 2021 and November 2021, along with the TESS observations from Sector~32 (2020), were used for the photometric LC modelling. The remaining data listed in Table~\ref{tab:phot_obs_summary} were used for the extraction of ToM. For ASAS~J082243+1927.0, the B, V, R, I, and C observations obtained between February 2022 and March 2022, along with the TESS observations from Sector~45 (2021), were used for LC modelling, while the remaining data listed in Table~\ref{tab:phot_obs_summary} were used for ToM extraction. Finally, for ZZ~PsA, the B, V, R, I, and C observations obtained between July 2021 and October 2021, along with the TESS observations from Sector~1 (25 July 2018 to 22 August 2018), were used for the LC modelling, and the remaining data listed in Table~\ref{tab:phot_obs_summary} were used for ToM extraction. 


\section{Spectroscopic analysis}\label{sec:spectr_anal}
\subsection{Data reduction}
Spectral data were reduced using the automated \texttt{EsoReflex}\footnote{\url{https://www.eso.org/sci/software/esoreflex/}} pipeline, which produced calibrated \texttt{.fits} frames. The pipeline applies all standard calibrations, including bias subtraction, flat-fielding, spectral extraction, and wavelength and flux calibrations. A custom Python script was developed to extract and visualise the flux–wavelength data. The flux values were retrieved from the \texttt{.fits} file headers and stored in tabular form. The start-exposure times provided in the headers were converted to mid-exposure times by accounting for the exposure duration, and subsequently transformed from modified Julian date (MJD) to heliocentric Julian date (HJD). The complete wavelength array was reconstructed using the starting wavelength and wavelength increment specified in the headers. 

\subsection{Broadening functions}\label{bf}
For each system, a set of synthetic spectra was selected from the POLLUX \footnote{\url{https://pollux.oreme.org}} database \citep{palacios2010, laverny2012}, to match the expected stellar parameters as closely as possible. Specifically, MARCS models from the AMBRE collection were deployed and optimised for F-G type stars, as the majority of low mass CBs consist of F, G, and K spectral type components \citep{step2012, rucinski2013}. This set was subsequently used for the broadening function (BF) computation (Section~\ref{bf}), radial velocity (RV) analysis (Section~\ref{rv}), and temperature estimation (Section~\ref{tempd}). The synthetic spectra were broadened to match the spectral resolution, and compared to the observed spectra within selected metal-line regions that are relatively insensitive to chromospheric activity, such as Fe~I, Ca~I, and Si~I absorption features \citep{rucinski2020}. The BFs were then derived through a custom optimisation-based approach, employing the differential evolution (DE) algorithm \citep{storn1997}. Differential evolution is a global, non-linear, optimisation method that iteratively evolves a population of candidate solutions to identify the optimal BF, by minimising the difference between the observed spectra and the synthetic spectra convolved with trial BFs. The DE approach has previously been successfully applied to map cyclotron emission regions on the surfaces of magnetic white dwarfs \citep{hakala2019,hakala2022}. The BF technique was preferred over the cross-correlation function (CCF) technique because it provides a cleaner separation of the component line profiles, leading to higher-quality RV profiles, particularly in systems with strong line blending and significant rotational broadening, such as CBs. In addition, the DE-based approach directly produces smooth BF profiles through the inclusion of a regularisation term in the optimisation process, eliminating the need for the additional post-processing smoothing commonly applied in the classical BF method \citep{rucinski2002b}.

\subsection{RV modelling and mass ratio}\label{rv}
The RV values of each spectrum were extracted from the best-fit BF models, by fitting two Gaussian functions (Equation~\ref{eq:gauss}), one for each stellar component, excluding outliers.
\begin{equation}
\label{eq:gauss}
f(x) = A_1 \cdot e^{\frac{-(x - B_1)^2}{2C_1^2}} + A_2 \cdot e^{\frac{-(x - B_2)^2}{2C_2^2}} + D
,\end{equation}
where the parameter $D$ was set to zero. The fitting function was iteratively evaluated and refined as needed to achieve the most accurate BF modelling parameters for both the primary and secondary components. The Gaussian centers $B_1$ and $B_2$, corresponding to the RVs of the primary ($V_1$) and secondary ($V_2$) components, were extracted for each of the 14 spectra. Barycentric corrections were applied to the extracted RVs using the values provided in the \texttt{.fits} file headers. The orbital phase of each spectrum was computed using the updated ephemerides and listed in Table~\ref{table_gen}. The orbital phase and the corrected RV values are listed in Table~\ref{rv_data}. The corresponding BF fits are presented in Figures~\ref{fig:allep_bf}, \ref{fig:asasj0822_bf}, and \ref{fig:zzpsa_bf} in the Appendix. No additional BF peak, indicating a third component, was detected in any of the spectra. Therefore, within the spectroscopic data sensitivity, there is no evidence for a third-light contribution in the systems.

\begin{figure}
\centering
\includegraphics[width=\columnwidth]{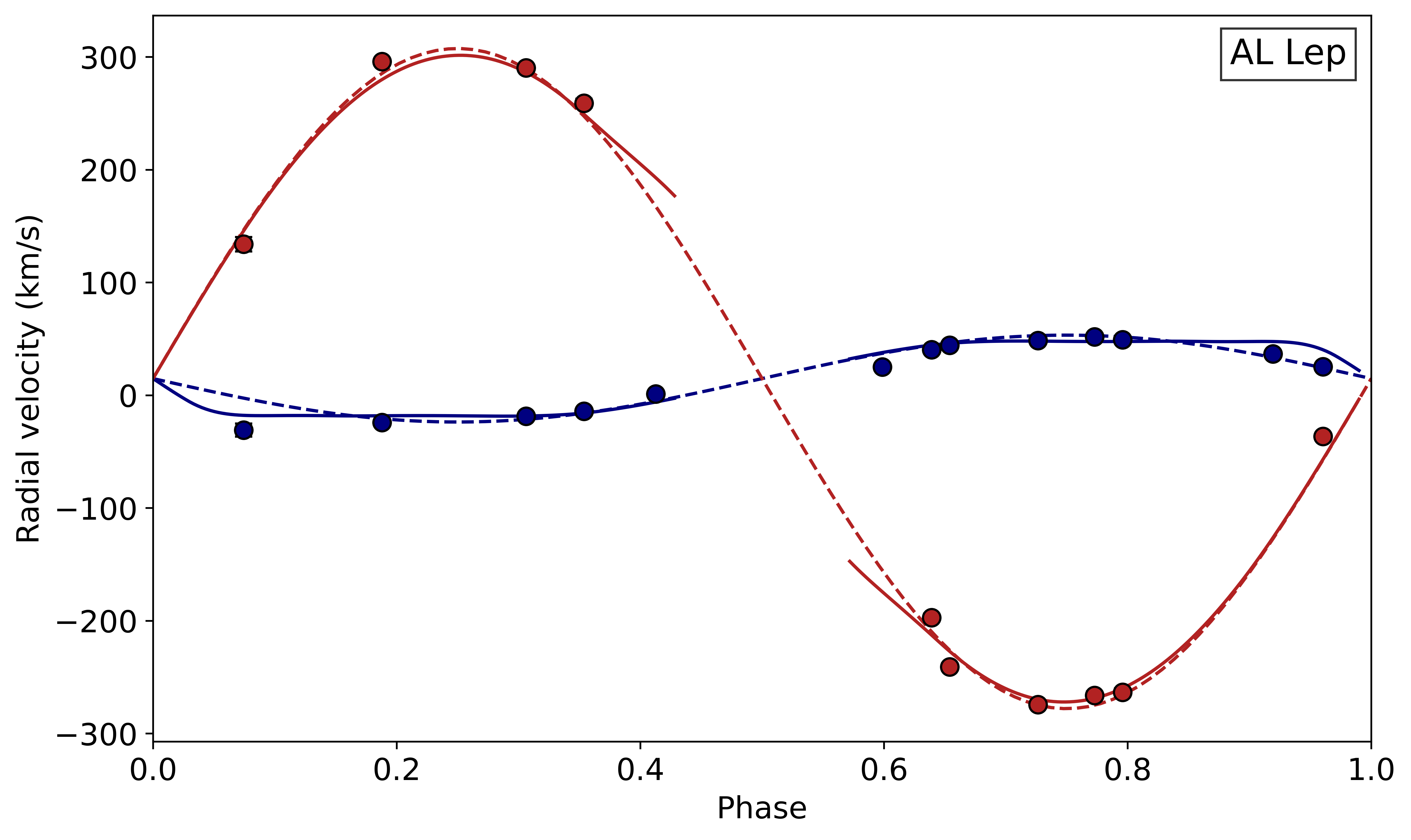}
\includegraphics[width=\columnwidth]{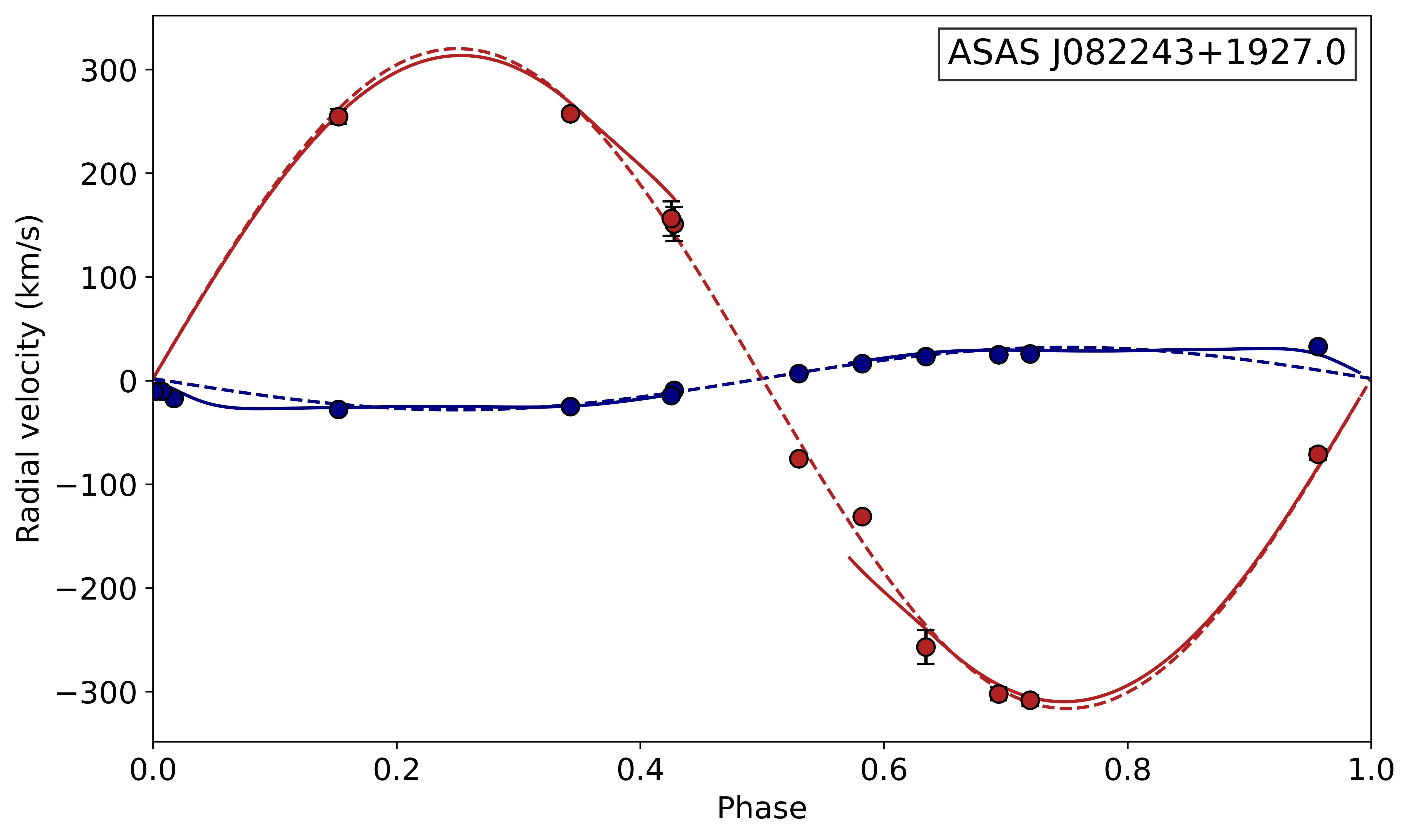}
\includegraphics[width=\columnwidth]{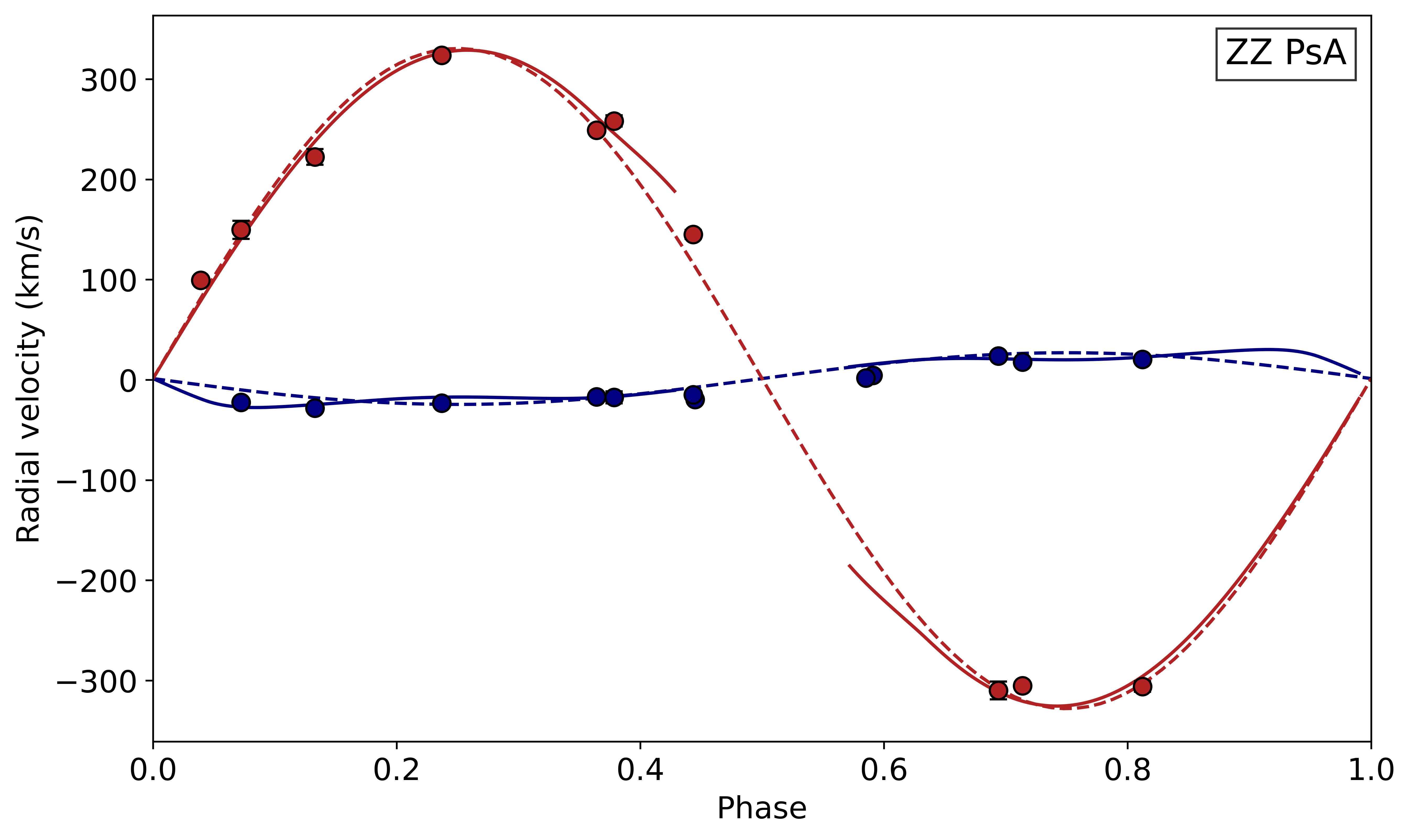}
\caption{RV curves for the three systems based on the extracted RVs. Dashed lines correspond to sinusoidal fits without tidal distortion corrections. Solid lines correspond to the RV curves corrected for proximity effects.}
\label{fig:rv_system}
\end{figure}

The resulting orbital phase and RV values were subsequently fitted using the sinusoidal functions of Equation~\ref{eq:rv}. The resulting parameters are presented in Table~\ref{tab:spectr_param}.
\begin{equation}\label{eq:rv}
V_{1,2}(\phi) = K_{1,2} \cdot \sin(2\pi\phi) + \gamma 
,\end{equation}
where $K_1$ and $K_2$ are the RV semi-amplitude values, $\phi$ is the orbital phase, and $\gamma$ is the systemic velocity. Data points near orbital phases 0.0 and 1.0 were excluded from the fitting procedure because of their increased uncertainties. Finally, the spectroscopic mass ratio, $q_{\rm sp}$, was derived from the relation $q_{\rm sp} = \vert K_1 \vert / \vert K_2 \vert$ (Table~\ref{tab:spectr_param}). Proximity effects in CBs, such as tidal distortion, mass transfer, and common-envelope interactions, can affect the observed RV curves and cause deviations from ideal sinusoidal profiles. These were taken into consideration through the MC modelling in Section~\ref{sec:pho_analysis}. The resulting RV-phase diagrams are shown in Figure~\ref{fig:rv_system}. The dashed curves represent the sinusoidal solutions, while the solid curves include the tidal distortion corrections from the MC model.

\subsection{Temperature and metallicity determination}\label{tempd}
The effective temperature and metallicity of the targets were determined by comparing observed spectra with synthetic spectra, spanning a range of temperature, metallicity, and surface gravity values \citep{rucinski2020}. MARCS models (AMBRE collection) synthetic spectra were deployed from the POLLUX database, as described in Section~\ref{bf}.

Late-type CBs typically present metallicity values ranging between $-0.65 \le [M/H] \le +0.5$ \citep{rucinski2013}. The model atmosphere grid adopted by \citet{rucinski2013} assumes a scaled-solar abundance; therefore, it was used here as an approximate reference interval for $[Fe/H]$. For F, G, and K-type CBs, the metallicity is expected to average around $\simeq -0.2$ \citep{rucinski2013}. Thus, the initial grid of synthetic spectra was assembled, assuming a slightly sub-solar iron abundance of $[Fe/H] = -0.25$, consistent with the old-disk nature of CBs and the available spectra in POLLUX database. Mild $\alpha$-enhancement models with $[\alpha/Fe] = +0.1$ were adopted, while models with $[\alpha/Fe] = +0.3$ were used to assess sensitivity to chemical composition. Considering uncertainties in the effective gravity of the tidally distorted components, surface gravity values were explored over the range $3.5 \le \log g \le 4.5$. 

Each synthetic spectrum was processed to reproduce the instrumental and physical broadening present in the observations. The synthetic resolving power was degraded to match that of the observed spectra via Gaussian convolution. Doppler shift was applied to account for orbital motion, and rotational broadening was modelled using the classical \citet{gray2005} kernel that incorporates linear limb darkening, under the assumption of synchronous rotation. The shape and width of the rotational broadening was determined by the projected rotational velocity, which was derived under the assumption of synchronous rotation, using the orbital parameters, spectroscopic velocity amplitude values, and the \citet{egg1983} Roche-lobe relation.

The comparison between the observed and synthetic spectra was performed using the line-depth ratio. More specifically, the depth of selected metal lines was normalised by the depth of a nearby Balmer line in both the observed and the reconstructed synthetic spectra. The best matching models were identified by minimising the total $\chi^2$ value, calculated from the agreement between the observed and synthetic line-depth ratio for all selected lines in each spectrum. This approach was preferred over direct calculation of the absolute line depth as it is less sensitive to local normalisation uncertainty and effects of rotational blending.

Magnetic heating in the stellar atmosphere produces a variety of chromospheric emission features, including Mg~I, Fe~I, Fe~II, Si~I, Si~II, and Ni~I, as well as hydrogen line series \citep{vernazza1981, linsky1980}. In W~UMa type CBs specifically, the rapid rotation and shared convective envelope drive enhanced chromospheric activity, and excess emission has been reported across multiple optical indicators \citep{rucinski1985, mitnyan2020, montes2000}. All metallic lines within the considered wavelength range were initially retrieved from the NIST database\footnote{\url{https://physics.nist.gov/PhysRefData/ASD/lines_form.html}} based on energy selection criteria, yielding a starting pool of 58 candidate lines. Following a quality assessment of line blending, signal-to-noise behaviour, and temperature sensitivity, this set was reduced to a final selection of 14 lines: 3 Mg~I lines (5167, 5173, 5184 \AA), 6 Fe~I lines (4891, 4919, 4957, 5267, 5328, 5370 \AA), 2 Fe~II lines (5167, 5206 \AA), 1 Si~I line (5006 \AA), 1 Si~II line (5041 \AA), and 1 Ni~I line (5082 \AA). The Mg~I triplet lines provide a reliable diagnostic of surface gravity and are among the strongest metallic features in the optical spectra of F-K type stars \citep{edvard1988, fuhrmann1997, osorio2015}. The Fe~I and Fe~II lines are particularly sensitive to both temperature and chromospheric heating across this wavelength window \citep{vieytes2024, hamann1992}, while the Si~II and Ni~I lines contribute additional temperature constraints \citep{vernazza1981}. The selected spectral lines, together with the best-fitting synthetic spectra and the corresponding observed spectra for each system, are shown in Figure~\ref{fig:lines}.

The effective temperature was determined by minimising the global mismatch metric, $\sum \chi^2$, computed over all selected spectral lines. The temperature dependence of the individual line contributions and of the total mismatch is shown in Figure~\ref{fig:chi2_teff}. Uncertainties were derived from the local curvature of the interpolated $\chi^{2}$ function. The highest-quality observed spectra, obtained with the UVES L arm and near-orbital phase 0.5, were adopted for this comparison, as line broadening at this phase is dominated by the primary component, ensuring that the derived temperature closely represents $T_{1}$. The resulting spectroscopic temperature values for the targets are presented in Table~\ref{tab:spectr_param}.

The metallicity of each system was determined after fixing the effective temperature to the value closest to the spectroscopic temperature obtained above. For each target, a metallicity grid of POLLUX synthetic spectra was constructed by varying $[Fe/H]$ and $[\alpha/Fe]$ while keeping $\log g$ fixed to the adopted value. The best model for each metallicity was selected by minimising over the available $[\alpha/Fe]$ values (Figure~\ref{fig:metal}). The synthetic spectra were processed in the same way as the temperature grid, including Doppler shift, rotational broadening, instrumental convolution, and interpolation onto the observed wavelength grid.


\section{Photometric analysis}\label{sec:pho_analysis}
All ToM were determined using the method of \citet{kwee1956}. From these timings, linear astronomical ephemerides were calculated and are listed in Table~\ref{table_gen}. The corresponding $O-C$ diagrams for each target in the sample were then constructed following the procedure described in Section~\ref{sec:oc}.

The LC analysis was performed using the WD code \citep{wilson1979, wils1990}, coupled with a Monte Carlo (MC) search algorithm, as described in detail by \citet{zola2004} and \citet{gazeas2021}. This approach does not require initial estimates for the free parameters, as it explores the entire parameter space within predefined ranges to identify the optimal solution.

The uncertainties of the fitted parameters were estimated through $\chi^{2}$ minimisation, following the methodology outlined in \citet{press1996}. The albedo and gravity darkening coefficients were fixed at their theoretical values of $A = 0.5$ and $g = 0.32$ \citep{lucy1967, rucinski1969}, appropriate for late-type stars with convective envelopes ($T < 6500$~K). Limb darkening coefficients were adopted from the tables of \citet{claret2011}, based on the effective temperature of the components and the photometric passband of each dataset.

The adjustable parameters include the orbital inclination ($i$), phase shift, effective temperature of the secondary component ($T_2$), dimensionless surface potentials ($\Omega_{1,2}$), luminosity of the primary component ($L_1$), and third light contribution ($l_3$). The luminosity of the secondary component ($L_2$) was not treated as a free parameter, as the IPB control parameter was set to zero. The temperature of the primary component ($T_1$) was considered fixed and equal to the spectroscopically determined ($T_{\rm eff}$). The spectroscopic mass ratio, $q_{\rm sp}$, was re-determined through the MC modelling as $q_{ph}$, in order to account for proximity effects. The resulting RV curves are shown as continuous lines in Figure~\ref{fig:rv_system}.

In the cases that the observed LCs exhibit asymmetries, such as unequal maxima (the O’Connell effect; \citealt{oco1951}), cool photospheric spots were introduced. The spot model includes four additional parameters describing its latitude, longitude, angular radius, and temperature factor. In all cases, a single spot on the primary component was sufficient to reproduce the observed asymmetry. 

\begin{figure}
\centering
\includegraphics[width=0.95\columnwidth]{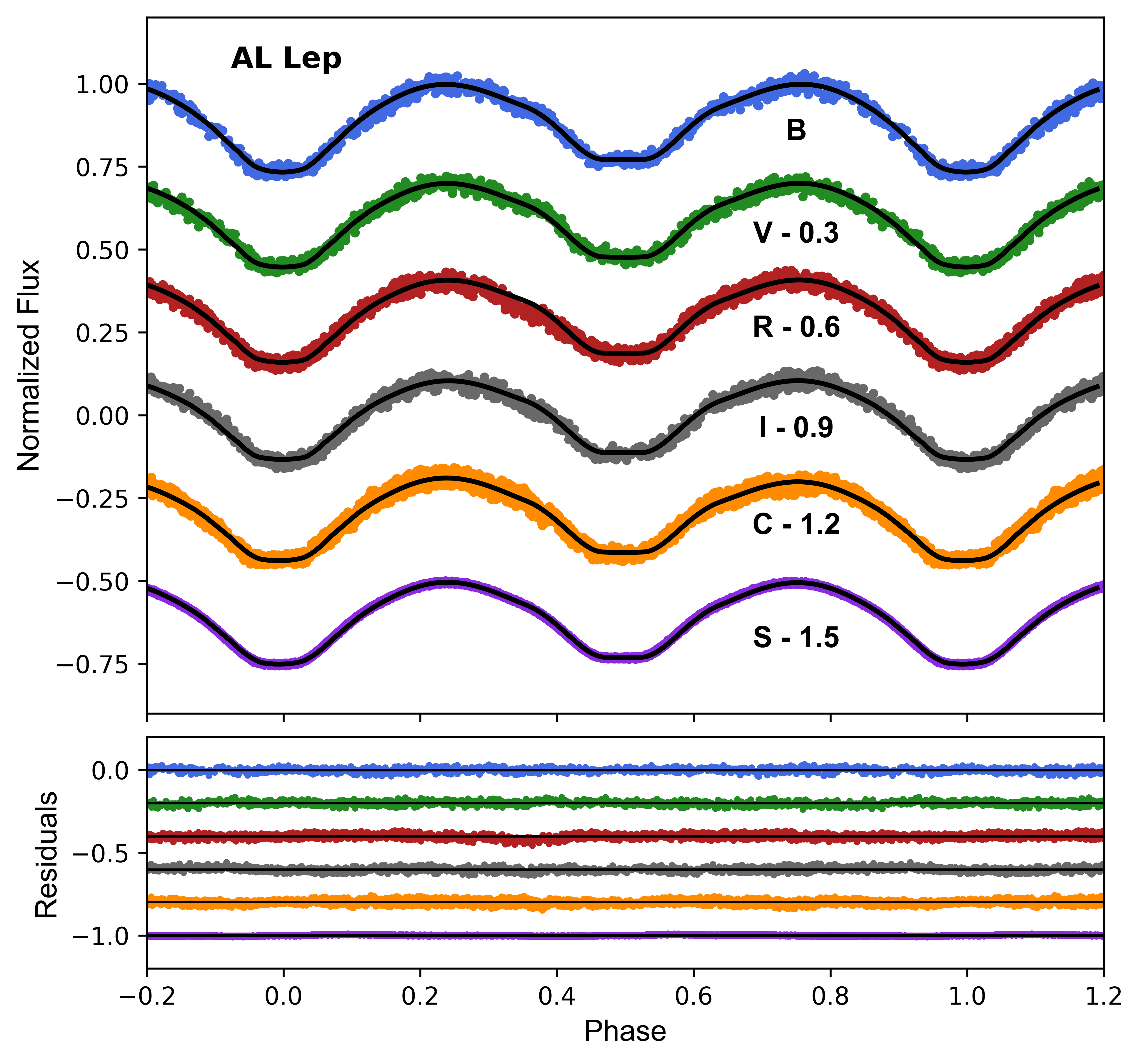}
\includegraphics[width=0.95\columnwidth]{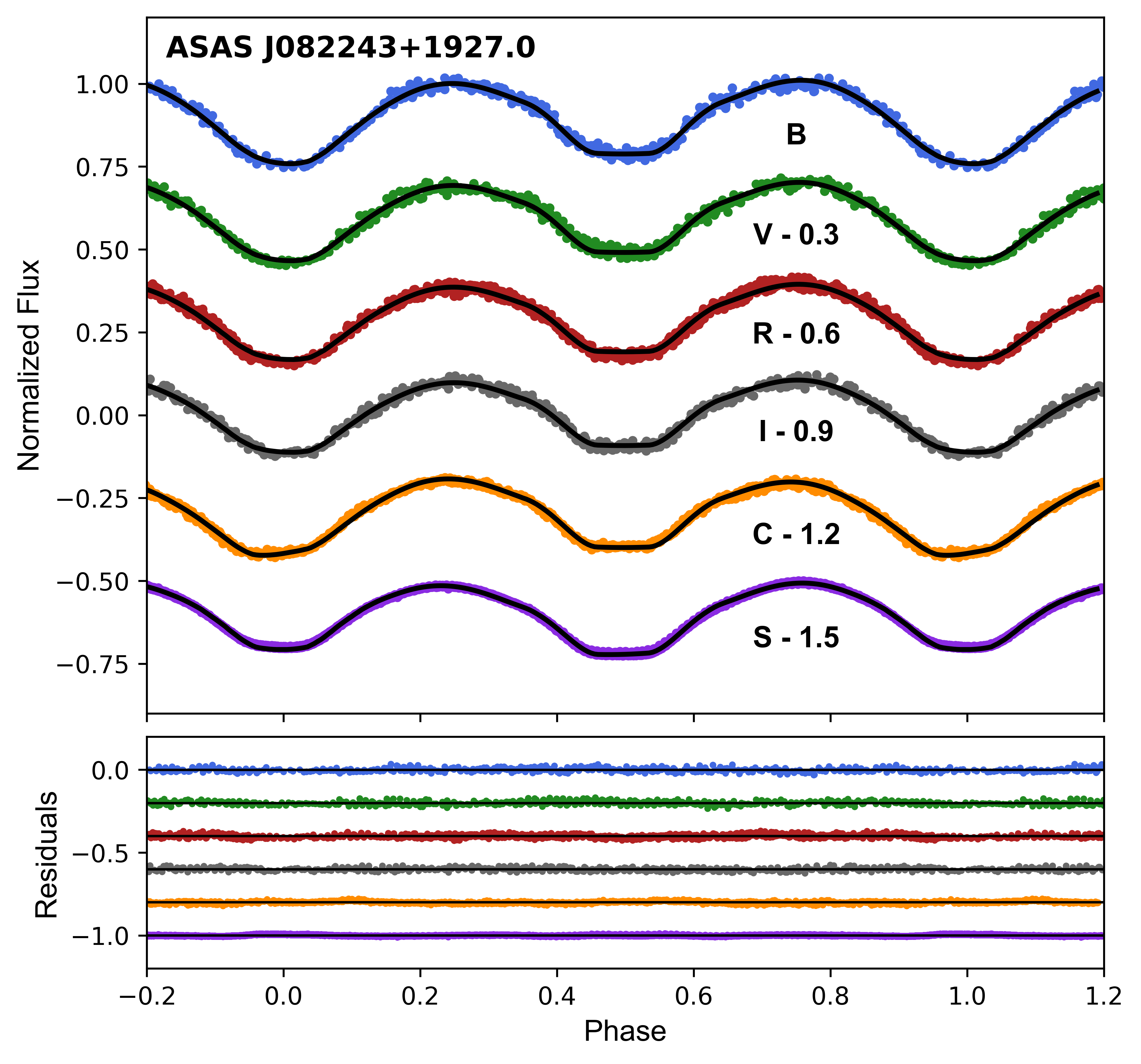}
\includegraphics[width=0.95\columnwidth]{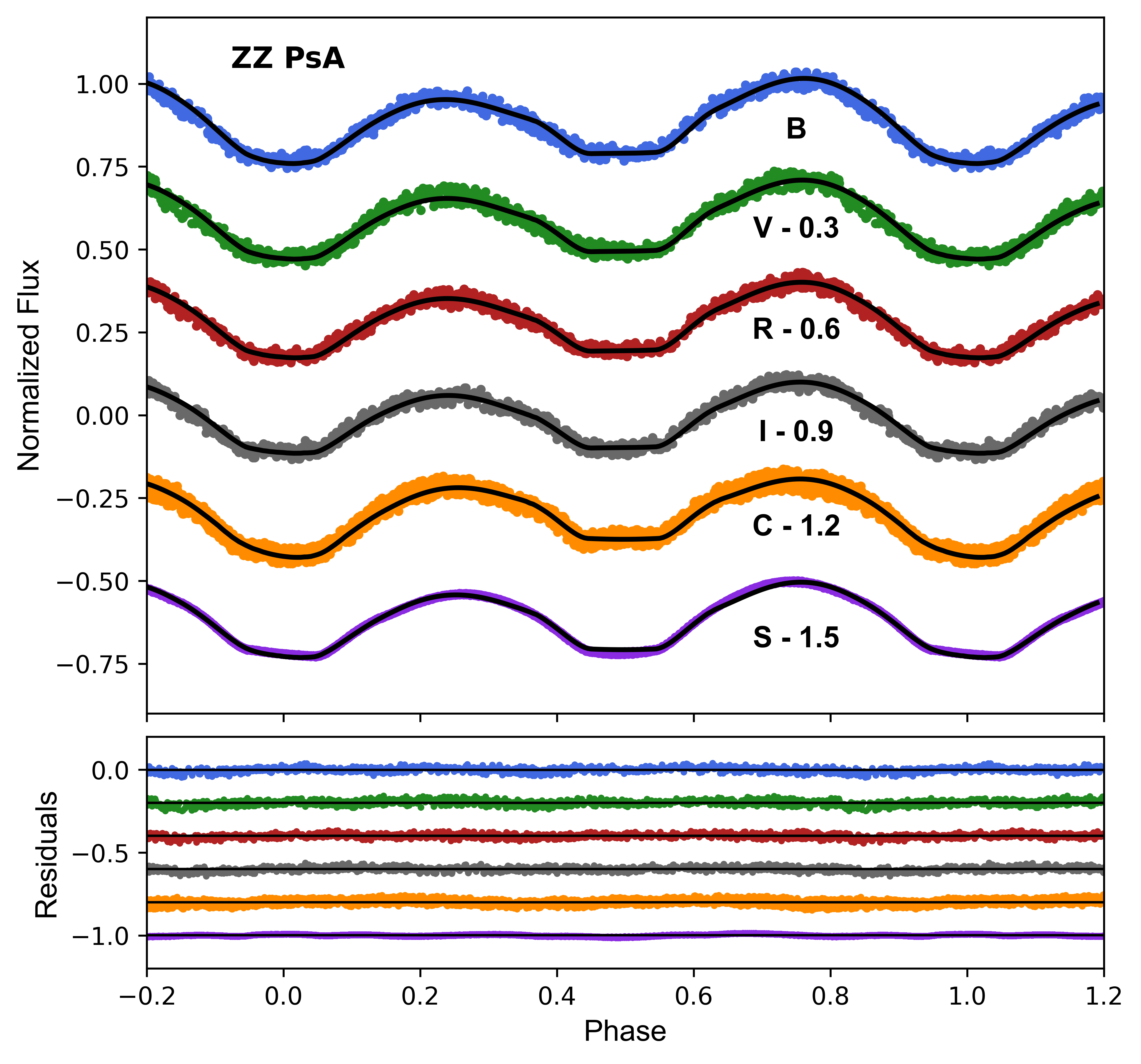}
\caption{Photometric data and adjusted photometric models of the systems AL~Lep, ASAS~J082243+1927.0 and ZZ~PsA. The models include the $B$, $V$, $R$, and $I$ filters following the Bessell specification, clear aperture ($C$), and the TESS passband ($S$).}
\label{fig:MC_system}
\end{figure}

The resulting models provide the physical and geometrical parameters of the systems, which are listed in Table~\ref{tab:model_parameters}, along with their $2\sigma$ uncertainties. The corresponding model fits to the observed LCs are presented in Figure~\ref{fig:MC_system}.

\section{O-C analysis}\label{sec:oc}
\begin{figure}[ht!]
\centering
\includegraphics[width=\columnwidth]{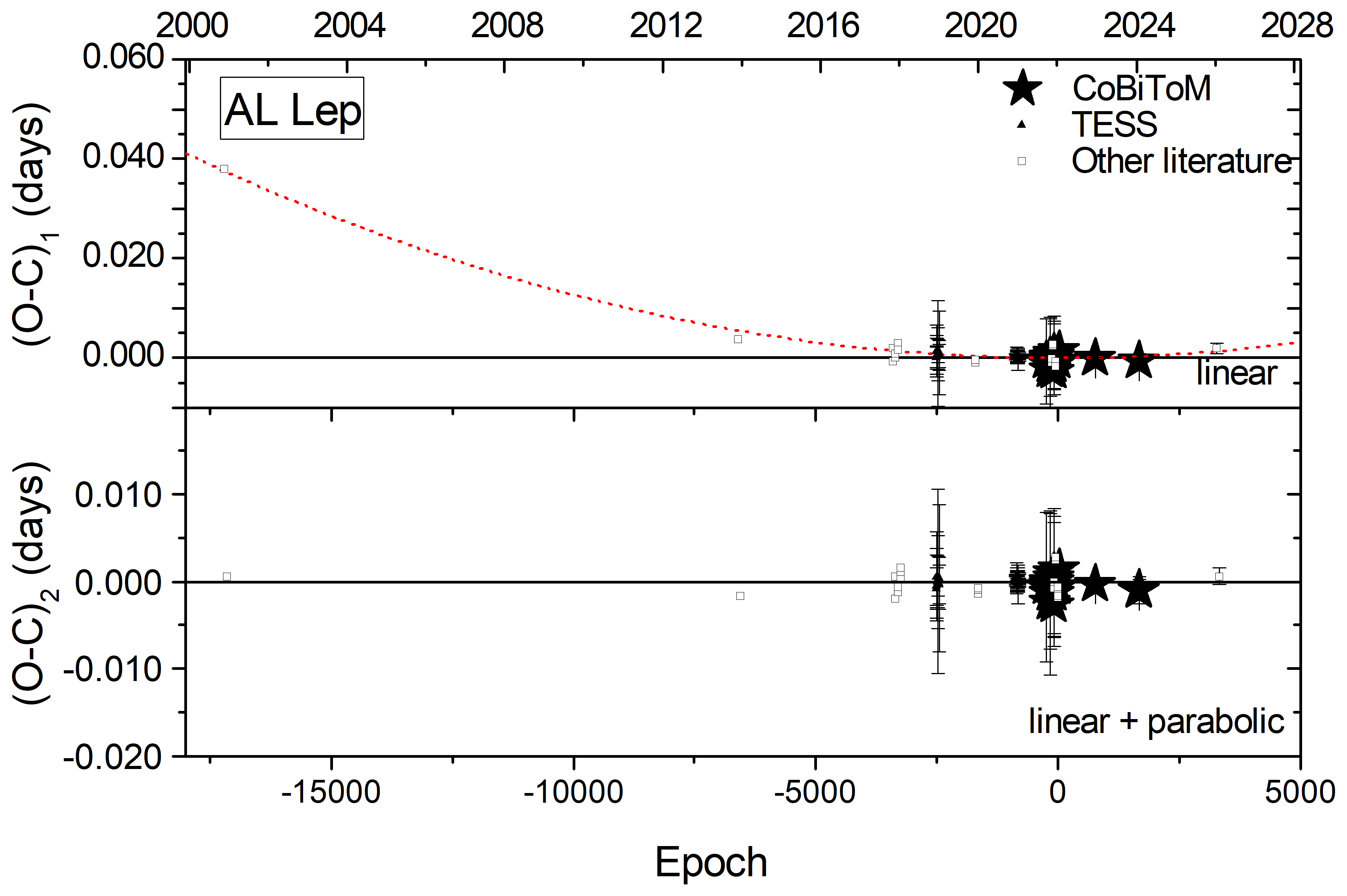}
\includegraphics[width=\columnwidth]{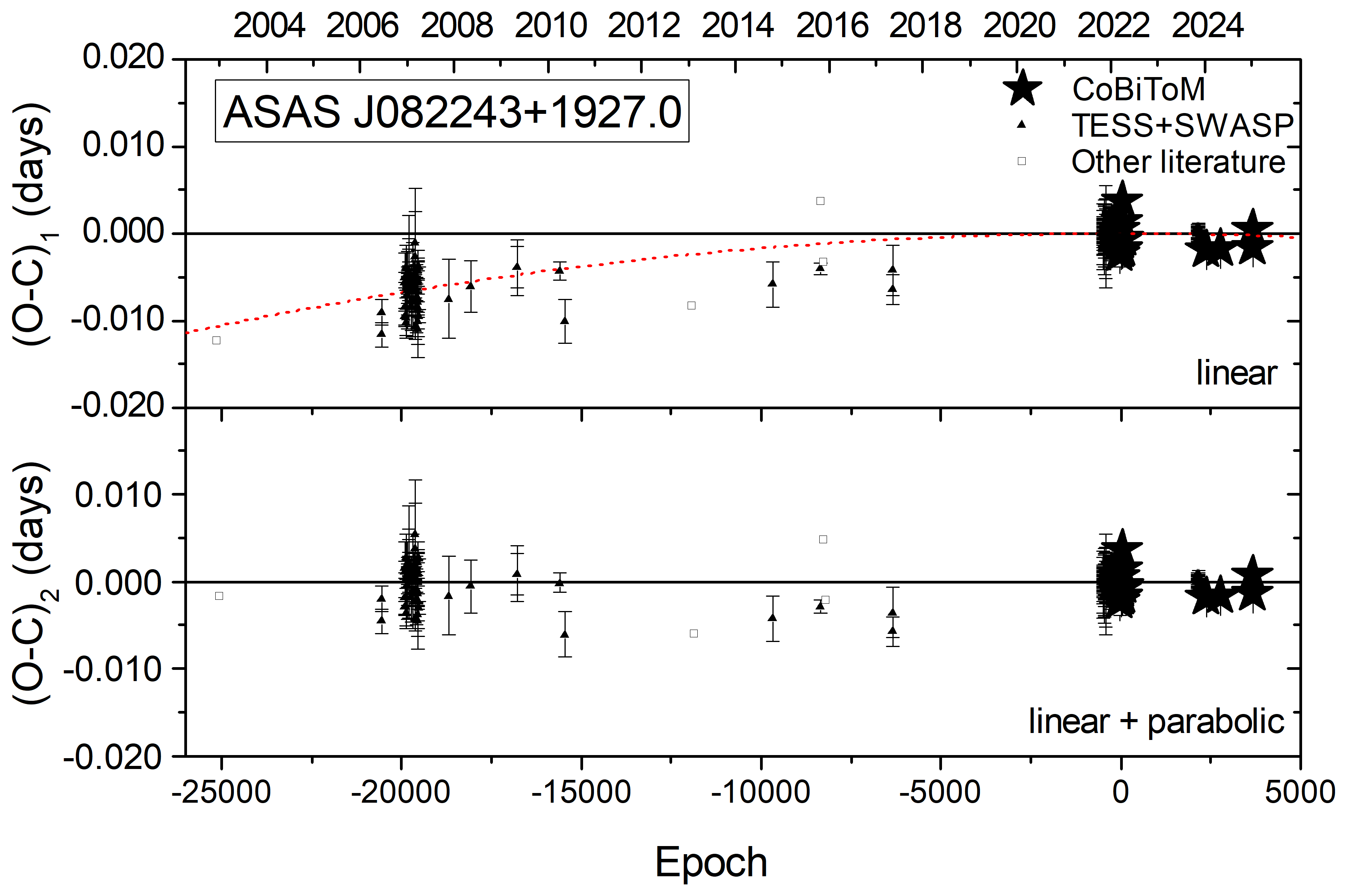}
\includegraphics[width=\columnwidth]{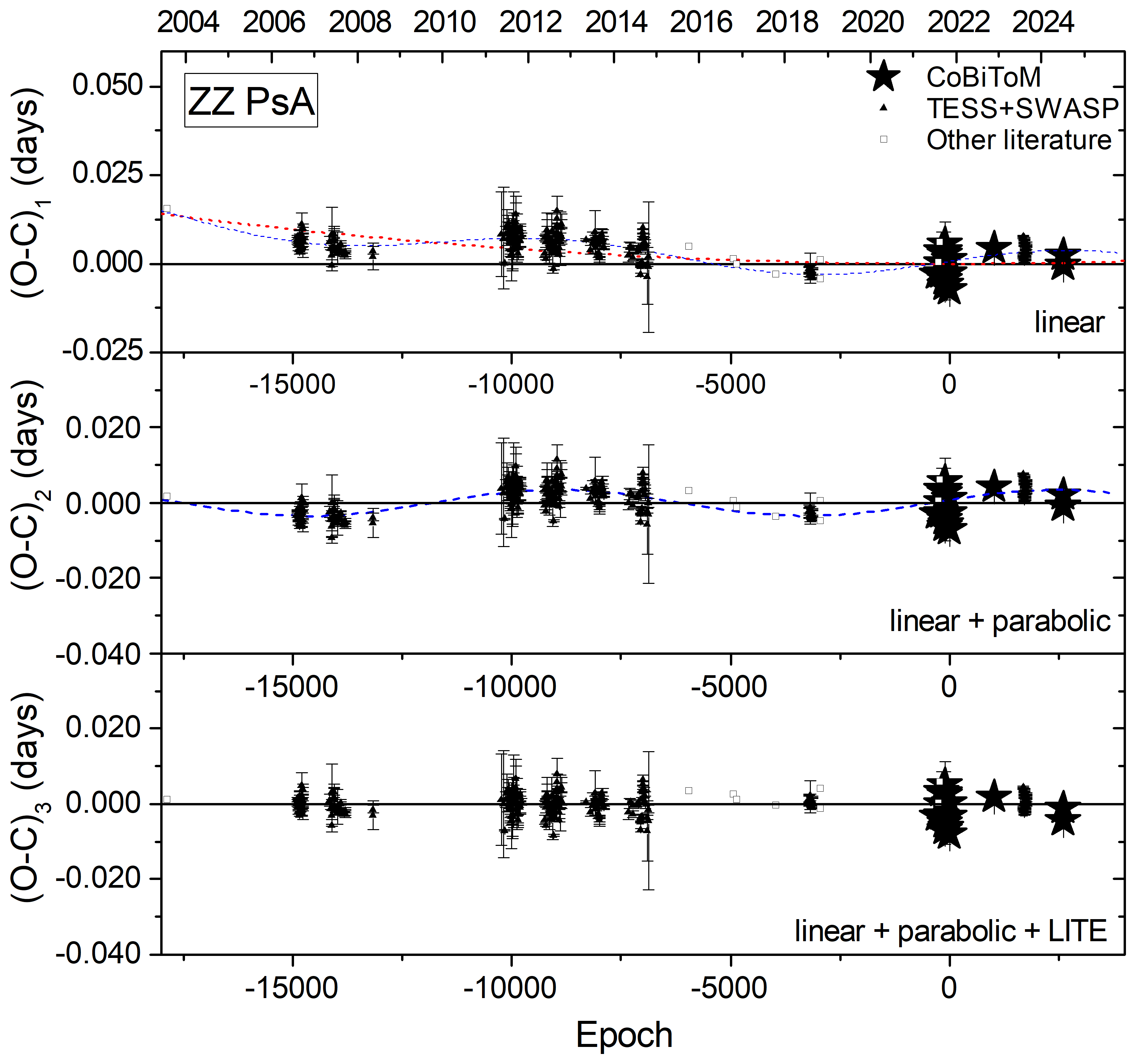}
\caption{$O-C$ diagrams for AL~Lep, ASAS~J082243+1927.0, and ZZ~PsA. The horizontal axes show the epoch number and corresponding calendar year, while the vertical axes show the $O-C$ residuals in days. $(O-C)_1$ corresponds to the residuals with respect to the linear ephemeris, $(O-C)_2$ to the residuals after including the parabolic term, and, for ZZ~PsA, $(O-C)_3$ to the residuals after including the cyclic term. The coloured curves represent the fitted terms, and different symbols denote the different sources of ToM.}
\label{fig:oc_diagrams}
\end{figure}

The orbital-period variations in the systems were investigated through the analysis of eclipse-timing residuals, constructed from the collected ToM. The resulting O-C diagrams for each system are shown in Figure~\ref{fig:oc_diagrams}. 

For AL~Lep and ASAS~J082243+1927.0, the $O-C$ diagrams show clear parabolic trends, indicating a secular variation of the orbital period. The residuals were therefore fitted with a quadratic ephemeris of the form of Equation~\ref{eq:oc1}.
\begin{equation}
\label{eq:oc1}
O-C = T_{min} - (T_0 + P E) - \frac{\beta}{2}E^2,
\end{equation}
where $T_{min}$ is the observed ToM light, $T_0$ and $P$ are the ephemeris elements (Table~\ref{table_gen}), $E$ is the cycle number, and $\beta$ is associated with the orbital-period variation. More specifically, the corresponding period change rate, expressed in $days/year$ units, is given as follows:
\begin{equation}
\dot{P} = \frac{dP}{dt} = \frac{\beta}{P}\times 365.25  
.\end{equation}
The upward parabola observed for AL~Lep corresponds to an increase in the orbital period, while the downward parabola found for ASAS~J082243+1927.0 indicates a period decrease. 

For ZZ~PsA, the $O-C$ diagram exhibits a cyclic variation, which may be caused either by a light-time effect due to an additional body \citep{irwin1952} or by magnetic activity in one or both components \citep{applegate1992, lanza2002}. However, no evidence of third light is found in either the photometric or spectroscopic data. Therefore, the cyclic modulation was modelled using a sinusoidal term, associated with magnetic activity in one or both components, as shown in Equation~\ref{eq:oc2} for zero eccentricity.

\begin{equation}
\label{eq:oc2}
O-C = T_{min} - (T_0 + P E) - \frac{\beta}{2}E^2 +
A \sin\left[\frac{2\pi\left(T_0 + PE - T_{mod} \right)}{365.25\,P_{\rm mod} }\right],
\end{equation}
where $A$ is the semi-amplitude of the cyclic modulation, $P_{\rm mod}$ is the modulation period expressed in years, and $T_{\rm mod}$ is the reference epoch of the cyclic modulation, expressed in $HJD$.

Given the contact configuration and late-type nature of the systems, the parabolic term can be considered either as a result of mass exchange between the components or non-conservative mass loss from the system. Assuming conservative mass transfer between the components \citep{hilditch2001}, the mass transfer rate was estimated from the relation:
\begin{equation}
\frac{\dot{P}}{P} = 3 \dot{M}_{T}\left(\frac{1}{M_{2}}-\frac{1}{M_{1}}\right) \Rightarrow \dot{M}_{T} = \frac{M_1 M_2}{3P(M_1-M_2)}\dot{P}
.\end{equation}
Positive values of $\dot{M}_{T}$ indicate mass transfer from the less massive secondary to the more massive primary, while negative values indicate mass transfer in the opposite direction.
In the case of the non-conservative scenario and assuming isotropic mass loss from the system, the mass loss rate can be estimated as
\begin{equation}
\frac{\dot{P}}{P} = -2\frac{\dot{M}_{\rm L}}{M_1+M_2} \Rightarrow \dot{M}_{\rm L} = -\frac{\dot{P}}{2P}(M_1+M_2)
\label{eq:mdot_loss_relation}
.\end{equation}
Following the non-conservative scenario of \citet{pribulla1999b}, in which mass is transferred between the components and lost from the system through the outer Lagrangian point $L_2$, carrying away orbital AM, the corresponding mass loss rate was estimated as
\begin{equation}
\frac{\dot{P}}{P} = \left[\frac{3(1+q)^2}{q}r(q)^2-3(1+q)+1 \right]\frac{\dot{M}_{\rm L,P}}{M_1+M_2},
\label{eq:mdot_l2_relation}
\end{equation}
or equivalently
\begin{equation}
\dot{M}_{\rm L,P}=\frac{M_1+M_2}{P}\frac{\dot{P}}{\left [\frac{3(1+q)^2}{q}r(q)^2-3(1+q)+1\right]},
\label{eq:mdot_l2}
\end{equation}
where $r(q)\simeq1.2$ is the distance of the $L_2$ point from the mass centre in units of the semi-major axis, and the spin AM is neglected.

In order to assess whether the resulted mass transfer rate is physically plausible, it was compared to the thermal-timescale mass transfer rate of the primary component. The thermal timescale, or else the Kelvin-Helmholtz timescale, represents the approximate time required for a star to radiate its gravitational binding energy and restore thermal equilibrium. The corresponding thermal timescale mass transfer rate was estimated as
\begin{equation}
\dot{M}_{\rm th} = \frac{M_1}{\tau_{\rm th,1}}
,\end{equation}
where $\tau_{\rm th,1} \simeq G M_1^2 / (R_1 L_1)$.
The agreement between the observed mass transfer rate and $\dot{M}_{\rm th}$ suggests that conservative mass exchange is a plausible explanation for the period variations. However, a significant difference may indicate that conservative mass transfer alone is not sufficient, and additional mechanisms, such as non-conservative mass loss, AML through magnetic braking, or outflow through the outer Lagrangian point, may also contribute. All results and obtained coefficients from the $O-C$ analysis of the three systems are listed in Table~\ref{tab:oc_coefficients}. As shown in Table~\ref{tab:oc_coefficients}, the absolute values of $\dot{M}_{\rm T}$ are lower than the corresponding $\dot{M}_{\rm th}$ values for all three systems, although the discrepancy is not large. Therefore, if the observed secular period variations are interpreted in terms of conservative mass transfer, the inferred mass exchange would proceed at a sub-thermal rate. In the case of ASAS~J082243+1927.0, where the secular period change is negative, additional AML or non-conservative processes may also contribute.

\begin{table*}
\caption{Fitted $O-C$ coefficients, period-change rate, mass transfer, and mass loss rate values for the three systems.}
\label{tab:oc_coefficients}
\centering
\begin{tabular}{lccc}
\hline\hline
Parameter & AL~Lep & ASAS~J082243 & ZZ~PsA \\
\hline
$\beta$ ($10^{-10}$~d cycle$^{-2}$) & $2.5364 \pm 0.0924$ 
& $-0.3372 \pm 0.0328$ & $0.8648 \pm 0.0002$ \\
$\dot{P}$ ($10^{-7}$~d yr$^{-1}$) & $2.0649 \pm 0.0752$ & $-0.4398 \pm 0.0428$ & $0.8448 \pm 0.0002$ \\
$A$ (d) & -- & -- & $0.0035 \pm 0.0891$ \\
$P_{\rm mod}$ (yr) & -- & -- & $11.7 \pm 0.4$ \\
$T_{\rm mod}$ (HJD) & -- & -- & $2455116.6 \pm 0.5$ \\
\hline
$\dot{M}_{\rm T}$ ($10^{-7}~M_{\odot}~{\rm yr}^{-1}$) & $0.39 \pm 0.13$ & $-0.07 \pm 0.04$ & $0.12 \pm 0.07$ \\
$\dot{M}_{\rm L}$ ($10^{-7}~M_{\odot}~{\rm yr}^{-1}$) & $-4.31 \pm 0.28$ & $1.10 \pm 0.13$ & $-2.19 \pm 0.15$ \\
$\dot{M}_{\rm L,P}$ ($10^{-7}~M_{\odot}~{\rm yr}^{-1}$) & $0.22 \pm 0.06$ & $-0.04 \pm 0.02$ & $0.07 \pm 0.04$ \\
$\tau_{\rm th}$ ($10^{7}~{\rm yr}$) & $1.22 \pm 0.16$ & $2.93 \pm 0.49$ & $1.92 \pm 0.29$ \\
$\dot{M}_{\rm th}$ ($10^{-7}~M_{\odot}~{\rm yr}^{-1}$) & $1.36 \pm 0.13$ & $0.44 \pm 0.05$ & $0.94 \pm 0.10$ \\
\hline
\end{tabular}
\end{table*}


\section{Tidal instability}
\begin{figure}[ht!]
\centering
\includegraphics[width=0.86\columnwidth]{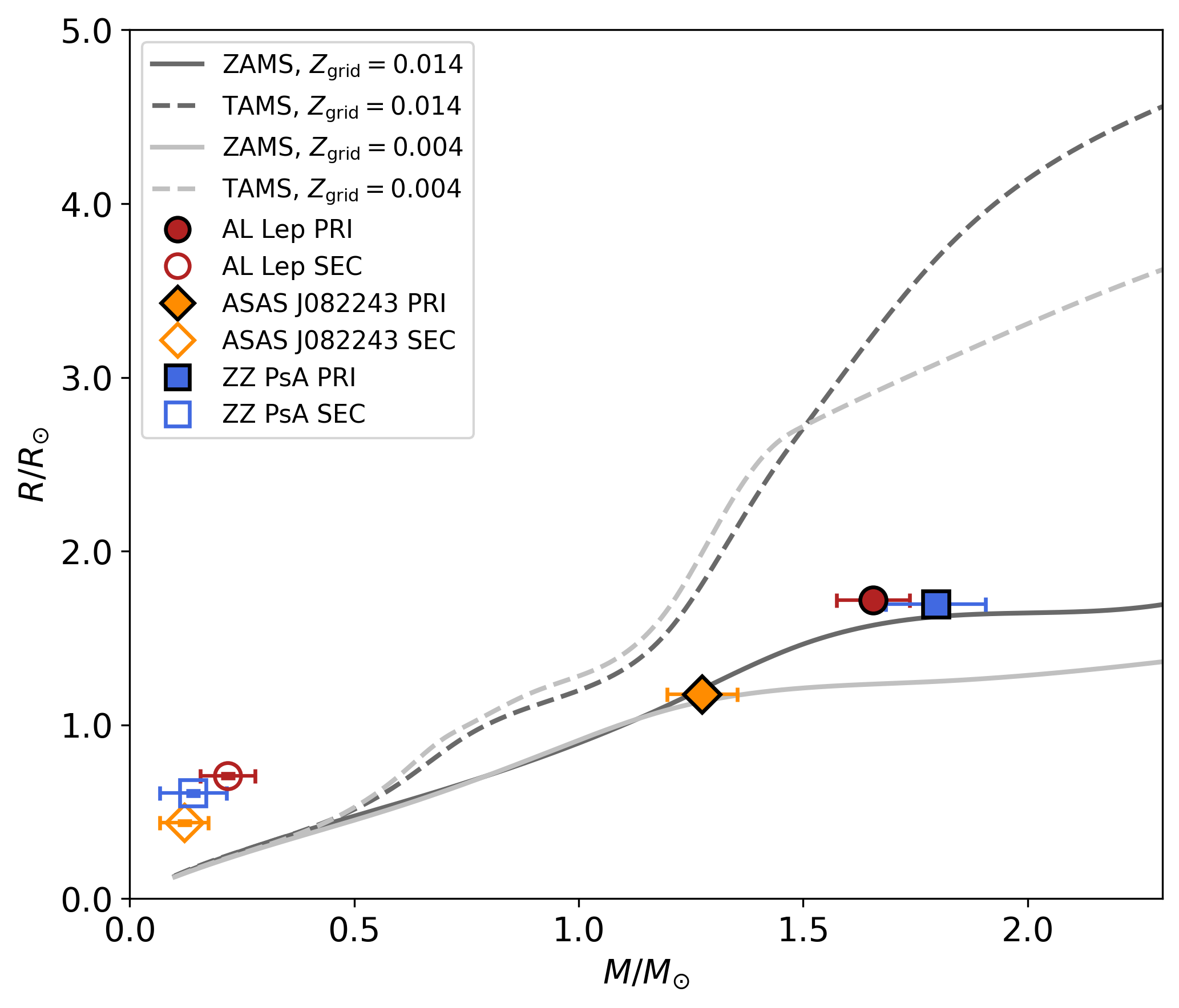}
\includegraphics[width=0.86\columnwidth]{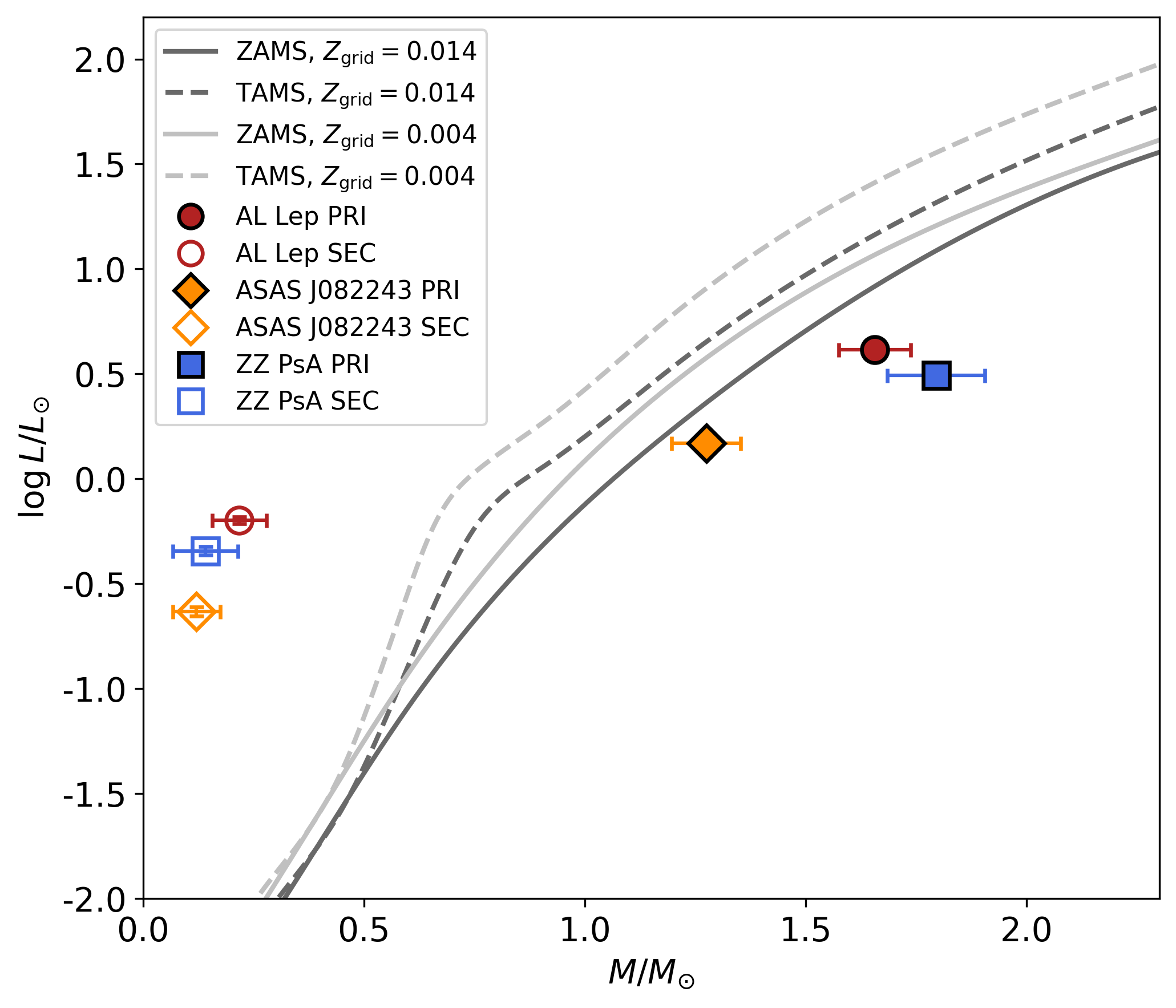}
\caption{Mass-radius ($M$-$R$) and mass-luminosity ($M$-$\log L$) diagrams for AL~Lep ($Z=0.01336$), ASAS~J082243+1927.0 ($Z=0.00455$), and ZZ~PsA ($Z=0.01380$). The location of the primary and secondary components is compared with PARSEC ZAMS and TAMS reference curves, computed for the metallicity grid closest to the spectroscopic value of each system.}
\label{fig:zams_tams}
\end{figure}

Physical parameters of CBs are highly constrained by their Roche geometry, and therefore they follow certain empirical relations \citep{maceroni1982, maceroni1983, hilditch1988, gazeas2006, gaz2008, gazeas2009}. The absolute parameters of both components were derived from the spectroscopic and photometric results described in the previous sections, and presented in Table~\ref{tab:abs_param}. The absolute parameters listed in Table~\ref{tab:abs_param} were compared with zero-age main sequence (ZAMS) and terminal-age main sequence (TAMS) reference curves, as shown in the mass-radius ($M$-$R$) and mass-luminosity ($M$-$\log L$) diagrams in Figure~\ref{fig:zams_tams}. The curves were derived from PARSEC evolutionary tracks, according to the metallicity of each system \citep{bressan2012}. The $M$-$R$ and $M$-$\log L$ diagrams indicate that the components of all three systems deviate from single-star main-sequence evolution, in accordance with the general properties of CBs discussed by \citet{gaz2008} and \citet{poro2026}. The primary components are relatively compact and under-luminous for their mass, while the secondary components are inflated and over-luminous compared to single MS stars of similar mass. 

\begin{table}[ht!]
\caption{Absolute parameters of AL~Lep, ASAS~J082243+1927.0, and ZZ~PsA.}
\label{tab:abs_param}
\centering
\begin{tabular}{lccc}
\hline\hline
Parameter & AL~Lep & ASAS~J082243 & ZZ~PsA \\
\hline
$M_{1}$ & 1.66 $\pm$ 0.08 & 1.28 $\pm$ 0.08 & 1.80 $\pm$ 0.11 \\
$M_{2}$ & 0.22 $\pm$ 0.06 & 0.12 $\pm$ 0.05 & 0.14 $\pm$ 0.07 \\
$R_{1}$ & 1.72 $\pm$ 0.03 & 1.18 $\pm$ 0.03 & 1.69 $\pm$ 0.04 \\
$R_{2}$ & 0.71 $\pm$ 0.01 & 0.44 $\pm$ 0.01 & 0.61 $\pm$ 0.02 \\
$L_{1}$ & 4.12 $\pm$ 0.33 & 1.48 $\pm$ 0.15 & 3.11 $\pm$ 0.25 \\
$L_{2}$ & 0.64 $\pm$ 0.02 & 0.23 $\pm$ 0.01 & 0.45 $\pm$ 0.02 \\
$M_{\rm bol,1}$ & 3.21 $\pm$ 0.47 & 4.32 $\pm$ 0.58 & 3.52 $\pm$ 0.46 \\
$M_{\rm bol,2}$ & 5.24 $\pm$ 0.22 & 6.33 $\pm$ 0.29 & 5.61 $\pm$ 0.28 \\
$a$ & 3.04 $\pm$ 0.06 & 2.01 $\pm$ 0.05 & 2.72 $\pm$ 0.07 \\
\hline
\end{tabular}
\tablefoot{Mass, radius, luminosity, and semi-major axis values are given in solar units. Bolometric magnitude values are given in mag.}
\end{table}

Theoretical modelling of low-mass MS stellar mergers remains uncertain, in terms of mass loss, AM redistribution, and the internal structure of the merger remnant \citep{mathieu2025, fabry2025}. As a result, the evolutionary status of a potential merger candidate cannot be reliably assessed by a single physical condition, and therefore the instability state of the three systems is examined through complementary criteria. As a first step the direct Darwin instability criterion was examined. The gyration radius of the primary star is a fundamental quantity in the Darwin instability criterion, since it determines the stellar moment of inertia through $I = k^2MR^2$. In previous studies, calibration relations between the primary gyration radius and the primary mass have been introduced \citep{arbutina2024,wadhwa2024}. However, these relations are not directly applicable to AL~Lep, ASAS~J082243+1927.0, and ZZ~PsA, because the mass and metallicity ranges covered by these calibrations do not fully include the physical parameters of the three systems. The primary gyration radius, $k_1$, was estimated from the stellar evolutionary models of \citet{claret2023}, who computed grids of stellar models for a wide range of mass, chemical composition, and internal-structure quantity values, including the gyration radius. For each of the three targets, the model grid was selected according to the metallicity closest to the spectroscopic $[Fe/H]$ determination, and the physical parameters of the systems. The gyration radius of the secondary is commonly fixed at $k_2^2 = 0.205$, which corresponds to a fully convective secondary approximated by an $n = 1.5$ polytropic configuration \citep{rasio1995, arbutina2024}.

\begin{figure}[ht!]
\centering
\includegraphics[width=\columnwidth]{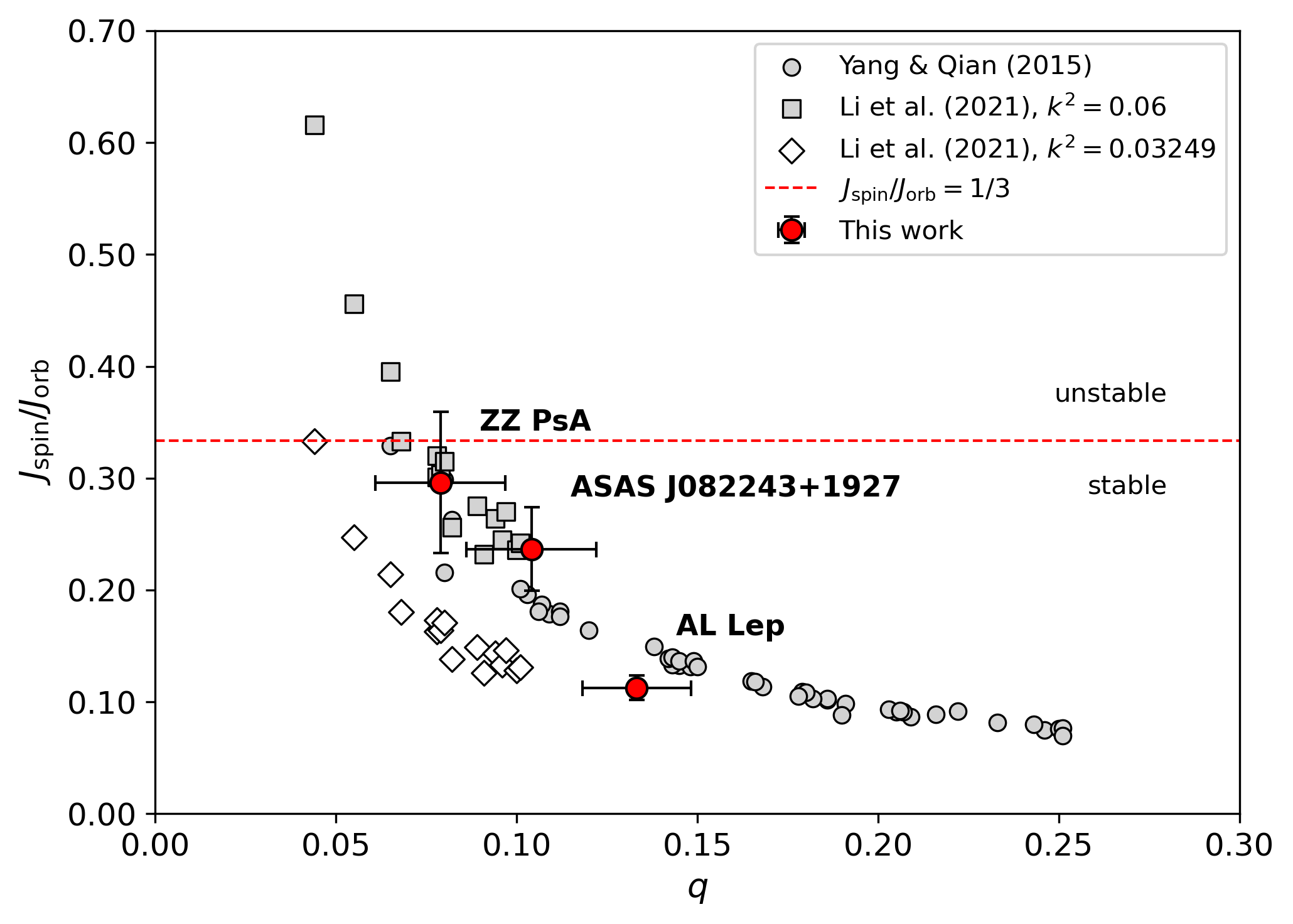}
\caption{Angular momentum ratio, $J_{\rm spin}/J_{\rm orb}$, as a function of the observed mass ratio, $q_{obs}$, for AL~Lep, ASAS~J082243+1927.0, and ZZ~PsA, denoted as red circles. Grey circles correspond to the sample of \citet{yang2015}, while grey squares and open diamonds to the original and revised values from \citet{li2021}. The dashed red line marks the Darwin instability limit.}
\label{fig:Jspin_Jorb_vs_q}
\end{figure}

Following \citet{pribulla1998}, the ratio of the spin and orbital AM, $J_{\rm spin}$ and $J_{\rm orb}$, is expressed as
\begin{equation}
\frac{J_{\rm spin}}{J_{\rm orb}} = \frac{1+q}{q}(k_1 r_1)^2 + (1+q)(k_2 r_2)^2
.\end{equation}
Here, $r_1$ and $r_2$ are the volume-equivalent fractional radius values extracted from the final MC models, which are equal to $r_{1,2} = R_{{\rm equiv},1,2}/a$, where $R_{{\rm equiv},1,2}$ are the volume-equivalent radius values of the primary and secondary components, and $a$ is the orbital separation. The resulting $J_{\rm spin}/J_{\rm orb}$ values for the three systems are given in Table~\ref{tab:q_inst} and plotted as a function of the observed mass ratio, $q_{obs}$, in Figure~\ref{fig:Jspin_Jorb_vs_q}. The three systems analysed in this work are shown as red circles, with their corresponding uncertainties. For comparison, the samples of \citet{yang2015} and \citet{li2021} are also shown. For \citet{li2021}, both the original values, calculated assuming $k^2 = 0.06$, and the revised values, obtained using $k^2 = 0.03249$, are presented. The original \citet{li2021} values above the Darwin instability limit illustrate the sensitivity of $J_{\rm spin}/J_{\rm orb}$ to the adopted gyration radius.

The instability mass ratio, $q_{\mathrm{inst}}$, was first determined for each system by numerically solving Eq.(13) of \citet{wadhwa2021}. The equation was solved by adopting fixed values of the fill-out factor, $f$, and the results are presented as $q_{\mathrm{inst,W}}$ in Table~\ref{tab:q_inst}. The boundaries $f=0$ and $f=1$ express marginal contact and full over-contact, respectively. As a second approach, $q_{\mathrm{inst}}$ was estimated using Eq.~(2) of \citet{jiang2026}. The obtained value of $q_{\min}$ (hereafter $q_{\mathrm{inst,J}}$) was then used to calculate the corresponding critical primary mass, $M'_{1,\mathrm{J}}$, from Eq.~(3) of \citet{jiang2026}. The derived values are listed in Table~\ref{tab:q_inst}.

\begin{table*}[ht!]
\caption{Gyration radius, instability mass ratio, and AMR of the systems.}
\label{tab:q_inst}
\centering
\begin{tabular}{lccc}
\hline\hline
Parameter & AL~Lep & ASAS~J082243 & ZZ~PsA \\
\hline
$k_{1}$                      & 0.192 $\pm$ 0.05 & 0.250 $\pm$ 0.07 & 0.232 $\pm$ 0.08 \\
$k_{1}^2$                    & 0.037 $\pm$ 0.02 & 0.062 $\pm$ 0.03 & 0.054 $\pm$ 0.04 \\
$q_{\rm inst,W}$ ($f=f_{\rm obs}$) & 0.055 $\pm$ 0.03 & 0.088 $\pm$ 0.05 & 0.079 $\pm$ 0.05 \\
$q_{\rm inst,W}$ ($0 \leq f \leq 1$) & 0.051 $\pm$ 0.02--0.056 $\pm$ 0.03 & 0.080 $\pm$ 0.04--0.091 $\pm$ 0.05 & 0.070 $\pm$ 0.04--0.079 $\pm$ 0.05 \\
$J_{\rm spin}/J_{\rm orb}$   & 0.113 $\pm$ 0.011 & 0.237 $\pm$ 0.037 & 0.296 $\pm$ 0.063 \\
$q_{\rm inst,J}$             & 0.050 $\pm$ 0.048 & 0.075 $\pm$ 0.059 & 0.050 $\pm$ 0.001 \\
$M'_{1,\rm J}$ ($M_{\odot}$) & 1.79 $\pm$ 0.85 & 1.31 $\pm$ 0.76 & 1.84 $\pm$ 0.12 \\
\hline
\end{tabular}
\end{table*}


\section{Discussion}
A system is considered dynamically unstable, and therefore a potential merger candidate, when it satisfies the Darwin instability criterion, $J_{\rm spin}/J_{\rm orb} > 1/3$, or when its observed mass ratio lies below the critical instability threshold, i.e. $q_{\rm obs} < q_{\mathrm{inst}}$. Additional indicators, such as a high fill-out factor and a strongly negative orbital-period derivative, may further support this classification. Beyond tidal instability, a system may also reach coalescence through thermal instability, which is associated with the advanced evolutionary state of the primary component, though it is a less common mechanism in W~UMa type CBs and particularly for low-mass stars \citep{jiang2026}. Regarding tidal instability, the \citet{wadhwa2021} formalism incorporates the internal structure through the gyration radius, as well as the degree of contact through the fill-out factor. On the other hand, the \citet{jiang2026} method is based on an approximate minimum mass ratio relation and the corresponding critical primary mass. Moreover, the physical parameters adopted in the sample are taken from \citet{latkovic2021}, in which several quantities are derived from empirical relations rather than from homogeneous spectroscopic and photometric modelling. Therefore, in this study, the \citet{wadhwa2021} formalism was adopted as the main mass ratio based criterion, while the \citet{jiang2026} method was only used as an evolutionary comparison.

AL~Lep lies well below the Darwin instability limit of $J_{\rm spin}/J_{\rm orb}=1/3$ (Table~\ref{tab:q_inst}, Figure~\ref{fig:Jspin_Jorb_vs_q}), indicating a dynamically stable system. This is also supported by the \citet{wadhwa2021} instability criterion, since the observed mass ratio remains significantly above the corresponding $q_{\mathrm{inst,W}}$. The \citet{jiang2026} criterion leads to the same conclusion, although this should be interpreted with caution due to the large uncertainty in $M'_{1,\mathrm{J}}$ (Table~\ref{tab:q_inst}). The O-C and LC analysis further support that AL~Lep is currently outside the tidal-instability regime, with a positive orbital period change rate of $\dot{P} = (2.0649 \pm 0.0752)\times10^{-7}~{\rm d~yr^{-1}}$ and a fill-out factor of $f=22\%$. Assuming conservative mass transfer, this implies mass transfer from the secondary to the primary component and further decrease of the mass ratio. Therefore, the system may be driven towards the LMR instability limit in the future. An approximate timescale for AL~Lep to reach the instability mass ratio can be estimated from
\begin{equation} 
\tau_{\rm inst} = \frac{M_2 - M_{2,\rm inst}}{\dot{M}_{\rm T}}=0.0031~{\rm Gyr}, 
\end{equation} 
where 
\begin{equation} M_{2,\rm inst} = \frac{q_{\rm inst}}{1+q_{\rm inst}} \left(M_1+M_2\right)=0.098~M_{\odot}. 
\end{equation}
In \citet{gaz2008} models, the time interval between the onset of contact and coalescence is typically of the order of $0.5-1.7~{\rm Gyr}$, depending on the initial component mass. Therefore, the estimated $\tau_{\rm inst}$ for AL~Lep is roughly two orders of magnitude shorter. Considering all the above criteria, the system is considered outside the tidal instability region, but in an advanced pre-coalescence stage. 

ASAS~J082243+1927.0 also remains below the Darwin instability threshold. However, its value of $J_{\rm spin}/J_{\rm orb}$ is much closer to the critical limit of $1/3$ than that of AL~Lep, indicating a smaller margin from the instability regime. The \citet{wadhwa2021} instability criterion also places the system outside the tidal-instability region, since the observed mass ratio is larger than the corresponding $q_{\mathrm{inst,W}}$ value. This is also supported by the \citet{jiang2026} criterion, which gives $M_1 < M'_{1,\mathrm{J}}$ (Table~\ref{tab:q_inst}). However, due to the small difference between $M_1$ and $M'_{1,\mathrm{J}}$, and the large uncertainty of $M'_{1,\mathrm{J}}$, this result should be interpreted with caution. The O-C analysis gives a negative orbital period change rate, $\dot{P} = (-4.3981 \pm 0.4284)\times10^{-8}~{\rm d~yr^{-1}}$, corresponding to a decreasing orbital period. Assuming a conservative mass transfer, this implies mass transfer from the more massive primary to the less massive secondary, an increase in the mass ratio, and therefore further deviation from the LMR instability limit. Thus, a future timescale to reach $q_{\rm inst}$ cannot be meaningfully estimated for ASAS~J082243+1927.0, unless a future reversal of the mass transfer is assumed. Although ASAS~J082243+1927.0 has a smaller margin from the Darwin instability threshold compared to AL~Lep, its relatively low fill-out factor of $25\%$ and its position below the Darwin threshold do not support its classification as a merger candidate.

ZZ~PsA presents a more ambiguous case. Its value of $J_{\rm spin}/J_{\rm orb} = 0.296 \pm 0.063$ lies slightly below the classical Darwin instability threshold of $1/3 (\simeq 0.333)$, suggesting stability. However, within the estimated uncertainties, its upper limit exceeds this critical value, indicating that the system may be consistent with an unstable configuration within the error limits. A similar classification was also suggested by \citet{wadhwa2021}, based on photometric modelling and the \citet{wadhwa2021} instability mass ratio criterion. The present study supports this classification using spectroscopically constrained parameters, since the nominal value of the observed mass ratio (Table~\ref{tab:model_parameters}) satisfies $q_{\rm obs} < q_{\mathrm{inst,W}}$ (Table~\ref{tab:q_inst}). However, the difference between $q_{\rm obs}$ and $q_{\mathrm{inst,W}}$ ($\simeq 0.0001)$ is much smaller than the uncertainties of both $q_{\mathrm{inst,W}}$ and $q_{\rm obs}$, indicating that this result should be interpreted with caution. According to the \citet{jiang2026} criterion, the system is not formally placed within the instability regime, since $M'_{1,\mathrm{J}} = 1.84$ is slightly higher than the observed primary mass, $M_1 = 1.80$ (Table~\ref{tab:q_inst}). Nevertheless, the small difference between these values indicates that ZZ~PsA lies close to the instability threshold according to this criterion as well. The $O-C$ analysis of ZZ~PsA reveals a positive orbital period change rate of $(8.4480 \pm 0.0019)\times10^{-8}~{\rm d~yr^{-1}}$, indicating an increasing orbital period. This indicates mass transfer from the less massive secondary to the more massive primary or mass loss through the outer Lagrangian point, $L_2$, considering the fill-out factor of $98\%$. This is consistent with strong energy and mass exchange through the common envelope \citep{fabry2025}. Since the system appears to be at, or slightly beyond, the adopted instability threshold, the remaining time needed to reach $q_{\rm inst}$ cannot be estimated as a future evolutionary timescale. Instead, only an upper-limit characteristic timescale can be defined as
\begin{equation} 
\tau_{\rm dep} = \frac{M_2}{|\dot{M}_{\rm T}|} = 0.0122 \pm 0.0098 ~{\rm Gyr}
\label{eq:tau_dep}
,\end{equation}
which assumes that the present mass transfer rate remains constant until the secondary mass is fully transferred. Comparing to \citet{gaz2008} models, the estimated $\tau_{\rm dep}$ for ZZ~PsA is roughly two orders of magnitude shorter, suggesting that the system may be in a very advanced pre-coalescence stage, especially if the evolutionary channel involved highly unequal initial component mass values and if coalescence mainly proceeds through mass transfer from the secondary to the primary. In practice, coalescence does not require the complete depletion of the secondary. The system may merge earlier because continued mass transfer and AM redistribution can lead to tidal instability before reaching the complete depletion phase. Considering its very high fill-out factor, the system may currently be in an extreme deep-contact phase of a TRO cycle \citep{qian2001,li2004}, or the common envelope may be close to unstable $L_2$ mass loss, rapid AML, and eventual coalescence \citet{webbink1976, rahunen1981,li2004}. According to the above, ZZ~PsA should be considered a borderline rather than an unambiguous merger candidate. The evolutionary status of ZZ~PsA remains uncertain and requires further long-term photometric and spectroscopic monitoring.


\section{Conclusions}
Overall, the three systems do not appear to have the same evolutionary state. ASAS~J082243+1927.0 seems to differ from AL~Lep and ZZ~PsA, both in terms of its O-C behaviour and its significantly lower metallicity. This may indicate that ASAS~J082243+1927.0 belongs to an older stellar population, while its present binary configuration suggests a less advanced CB evolutionary stage compared to the other two systems. In contrast, AL~Lep and ZZ~PsA show increasing orbital periods, which may indicate that they have already entered a later phase of their contact evolution. This is qualitatively consistent with the evolutionary models of \citet{gaz2008}. In this study, their model calculations show that systems with a lower total mass evolve more slowly and reach coalescence several gigayears later, while systems with higher initial total mass evolve faster. Moreover, systems with a larger initial difference between the component mass values tend to reach coalescence earlier than systems with more comparable initial mass values. In this context, the lower total mass of ASAS~J082243+1927.0 may correspond to a slower evolutionary path, supporting the view that it is less evolved than AL~Lep and ZZ~PsA. Also, according to the models presented in Table~3 of \citet{gaz2008}, the orbital period at coalescence is larger than the period at the start of contact for systems with lower $q$ and higher $M_1$ values. This suggests that an orbital-period increase may occur during the later stages of contact evolution, prior to coalescence, especially for LMR CBs. Such a behaviour is observed for AL~Lep and ZZ~PsA, whose O-C diagrams show positive quadratic terms. On the other hand, ASAS~J082243+1927.0 shows a negative quadratic term, indicating a decreasing orbital period. This may imply that the system is still in an earlier evolutionary phase, before the possible turning point towards an orbital-period increase. This is shown in Fig.~5 of \citet{gaz2008}, where the evolutionary tracks indicate that CBs may undergo a phase of decreasing orbital period before reaching a turning point and evolving towards larger periods as they approach coalescence. Therefore, ASAS~J082243+1927.0 may represent a lower-mass and less evolved system that is still in the period-decrease phase, whereas AL~Lep and ZZ~PsA may have already progressed to a later stage characterised by orbital-period increase. This is also in agreement with the correlations reported by \citet{qian2003}, which suggest that systems with more massive primary components tend to show a larger orbital period change rate, as AL~Lep and ZZ~PsA display higher $M_1$ values compared to ASAS~J082243+1927.0. Nevertheless, the above interpretation should be regarded as qualitative, since the evolutionary tracks of \citet{gaz2008} and \citet{qian2003} are not tailored to the exact mass, metallicity, and mass transfer history of the systems in this study.


\section{Data availability}
The reduced UVES spectra are available through the ESO Phase 3 archive (\url{https://archive.eso.org/scienceportal/home}, DOI: 10.18727/archive/50) under ESO programmes 112.25W8.001, 112.25W8.002, and 112.25W8.003. The NOA photometric data and derived ToM are provided as ASCII tables at Zenodo (\url{https://zenodo.org/}, DOI: 10.5281/zenodo.21625711) and CoBiToM project website (\url{http://users.uoa.gr/~kgaze/cobitom/publications.html}).

\begin{acknowledgements}
Based on observations collected at the European Southern Observatory under ESO programme(s) 112.25W8.001, 112.25W8.002, and 112.25W8.003. These observations were obtained as part of the programme "Merging Binaries -- Identifying the Candidates for Red Nova Events" (PI: P.~Hakala; Co-Is: K.~Gazeas and S.~Palafouta). Part of the photometric observations used in this work were obtained with the 1.2~m Kryoneri telescope, operated by the Institute for Astronomy, Astrophysics, Space Applications and Remote Sensing of the National Observatory of Athens, located in Corinthia, Greece. These observations were supported by the Europlanet 2024 Research Infrastructure (RI) through the NA2 programme. Europlanet 2024 RI received funding from the European Union’s Horizon 2020 research and innovation programme under grant agreement No.~871149. The rest of the photometric data were obtained with the 0.4~m robotic telescope of the University of Athens Observatory (UOAO), located at the National and Kapodistrian University of Athens, Greece \citep{gaz2016}. The authors wish to thank the staff and collaborators at the above observing facilities for their support and for the telescope time allocation.
\end{acknowledgements}

\bibliographystyle{aa}
\bibliography{aa61420-26.bib}

\begin{appendix}
\onecolumn
\FloatBarrier

\section{Observational data and observing logs}

\begin{table}[ht!]
\caption{Spectroscopic runs and data acquisition details.}
\label{vlt_appl}
\centering
\begin{tabular}{lccc}
\hline\hline
System          & AL~Lep       & ZZ~PsA        & ASAS~J082243+1927.0 \\
\hline
Instrument      & UVES          & UVES          & UVES \\
Telescope setup & UT2           & UT2           & UT2 \\
Mode            & SM            & SM            & SM \\
Exposure time   & 300~s         & 300~s         & 300~s \\
Total time      & 04h~40m       & 04h~40m       & 06h~20m \\
Date            & 2023 Nov. 2 and 2023 Dec. 18 & 2023 Nov. 2 & 2024 Jan. 23 \\
S/N (L, U arm)  & L: $\sim$59-188, U: $\sim$70-212 & L: $\sim$55-114, U: $\sim$68-136 & L: $\sim$72-205, U: $\sim$87-240 \\
\hline
\end{tabular}
\end{table}

\begin{table}[ht!]
\caption{Summary of the photometric observations used for the light-curve analysis and times-of-minimum determination.}
\label{tab:phot_obs_summary}
\centering
\begin{tabular}{lllll}
\hline\hline
System & Filter/passband & Observing period & Exposure time & Data source \\
\hline
AL~Lep & $B$ & Aug. 2021--Nov. 2021 & 90~s & UOAO \\
       & $V$ & Aug. 2021--Nov. 2021 & 40~s & UOAO \\
       & $R$ & Aug. 2021--Jan. 2024 & 20~s, 30~s & UOAO \\
       & $I$ & Aug. 2021--Nov. 2022 & 20~s, 30~s & UOAO \\
       & Clear aperture ($C$) & Dec. 2021--Jan. 2022 & 10~s & UOAO \\
       & TESS ($S$) & Sectors 5 and 32 (2018, 2020) & 600~s, 1800~s & TESS-SPOC \\
\hline
ASAS~J082243+1927.0 & $B$ & Feb. 2022--Mar. 2022 & 90~s & Kryoneri Obs. \\
                    & $V$ & Feb. 2022--Mar. 2022 & 60~s & Kryoneri Obs. \\
                    & $R$ & Feb. 2022--Dec. 2023 & 30~s & Kryoneri Obs. \\
                    & $I$ & Feb. 2022--Mar. 2022 & 30~s & Kryoneri Obs. \\
                    & Clear aperture ($C$) & Mar. 2022 & 20~s & UOAO \\
                    & SWASP & Sep. 2004--Mar. 2011 & 30~s & SWASP\\
                    & TESS ($S$) & Sectors 44--46, 71, and 72 (2021--2023) & 200~s, 600~s & TESS-SPOC \\
\hline
ZZ~PsA & $B$ & Jul. 2021--Oct. 2021 & 90~s & UOAO \\
       & $V$ & Jul. 2021--Oct. 2021 & 40~s & UOAO \\
       & $R$ & Jul. 2021--Jul. 2024 & 20~s, 30~s & UOAO \\
       & $I$ & Jul. 2021--Oct. 2021 & 20~s & UOAO \\
       & Clear aperture ($C$) & Nov. 2021 & 15~s & UOAO \\
       & SWASP & Aug.2006--Nov. 2014 & 30~s & SWASP\\
       & TESS ($S$) & Sectors 1 and 68 (2018--2023) & 200~s, 1800~s & TESS-SPOC \\
\hline
\end{tabular}
\end{table}

\FloatBarrier

\section{Spectroscopic analysis and radial velocity measurements}

\begin{figure}
\centering
\includegraphics[width=0.22\textwidth]{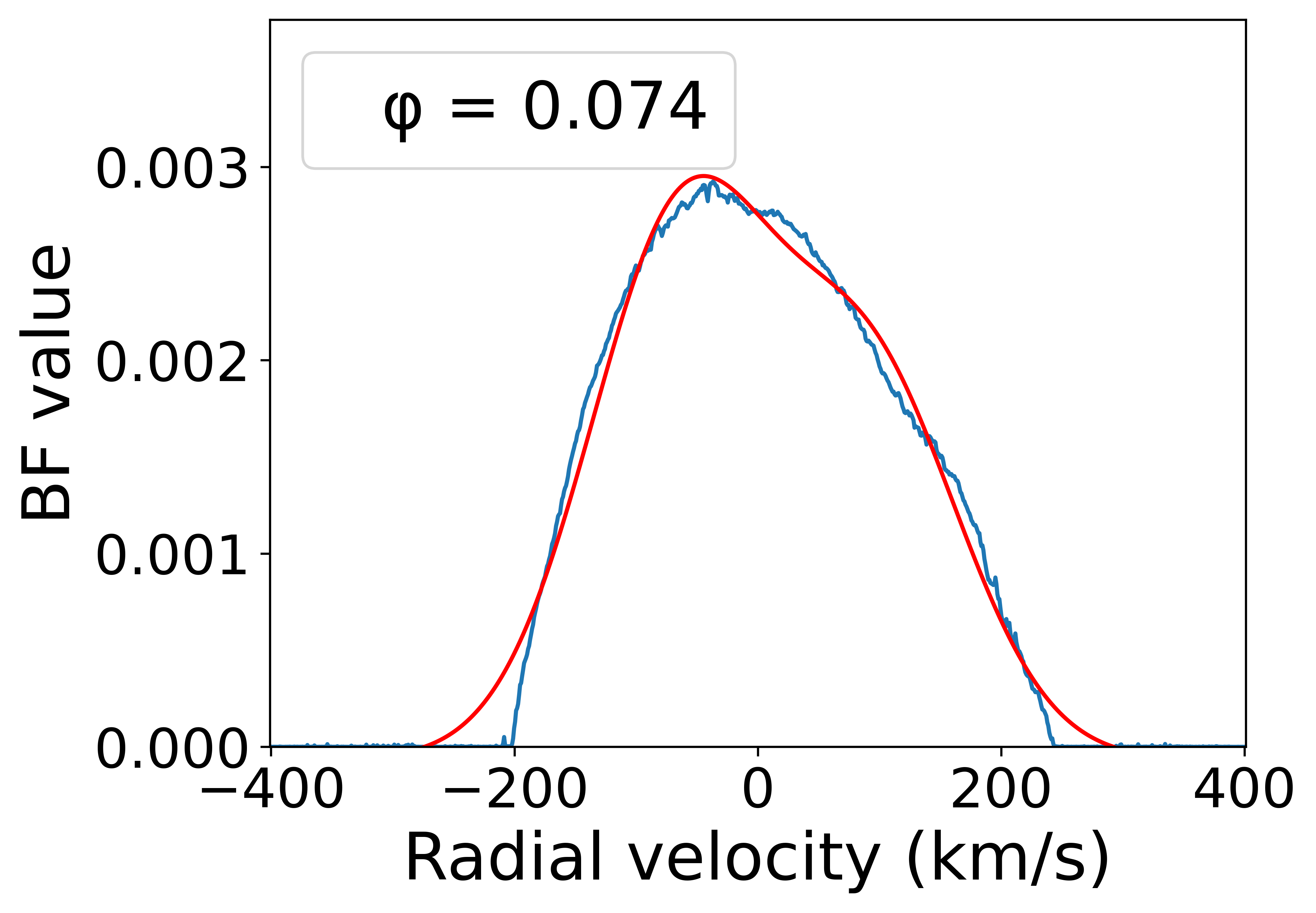}
\includegraphics[width=0.22\textwidth]{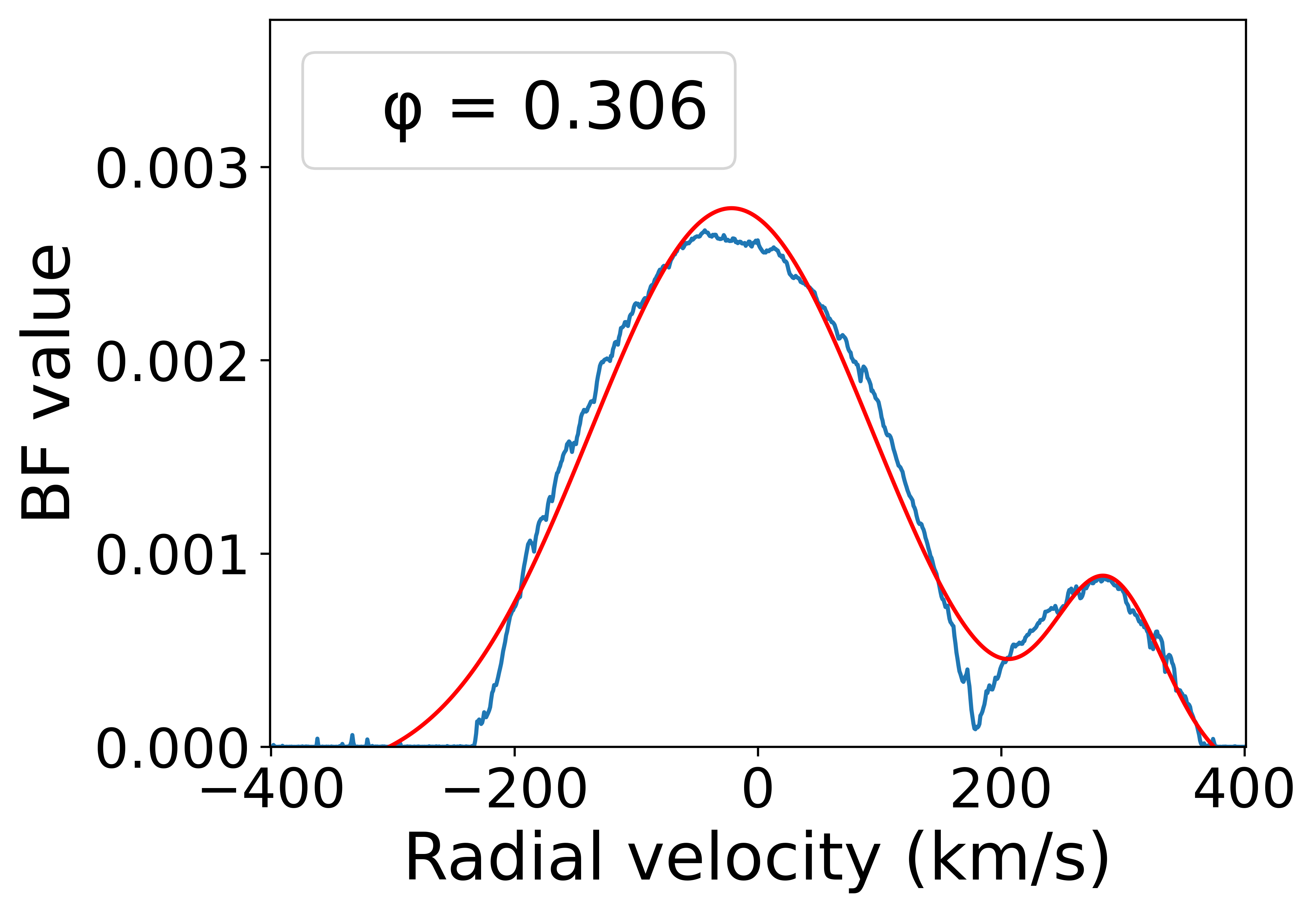}
\includegraphics[width=0.22\textwidth]{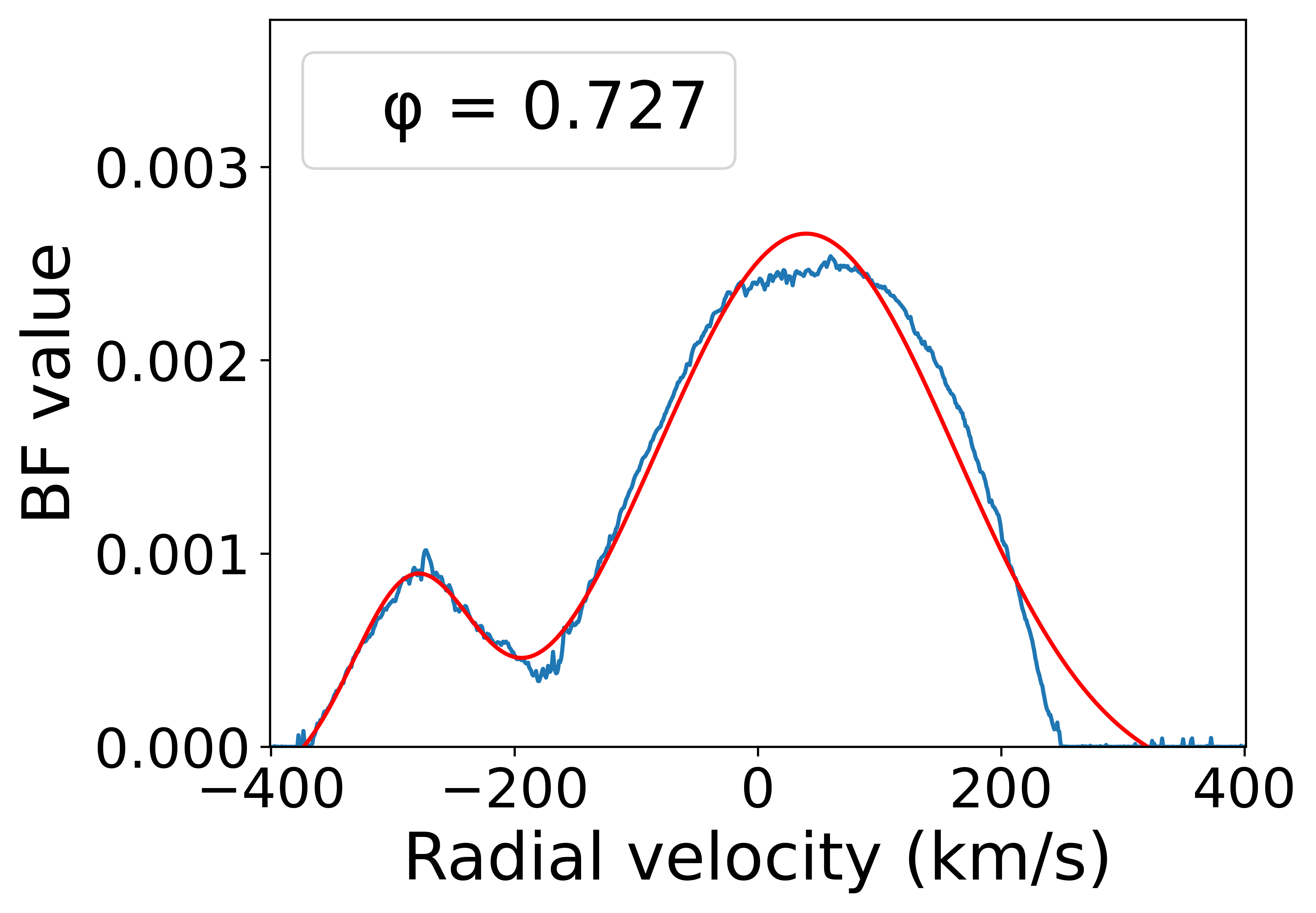}
\includegraphics[width=0.22\textwidth]{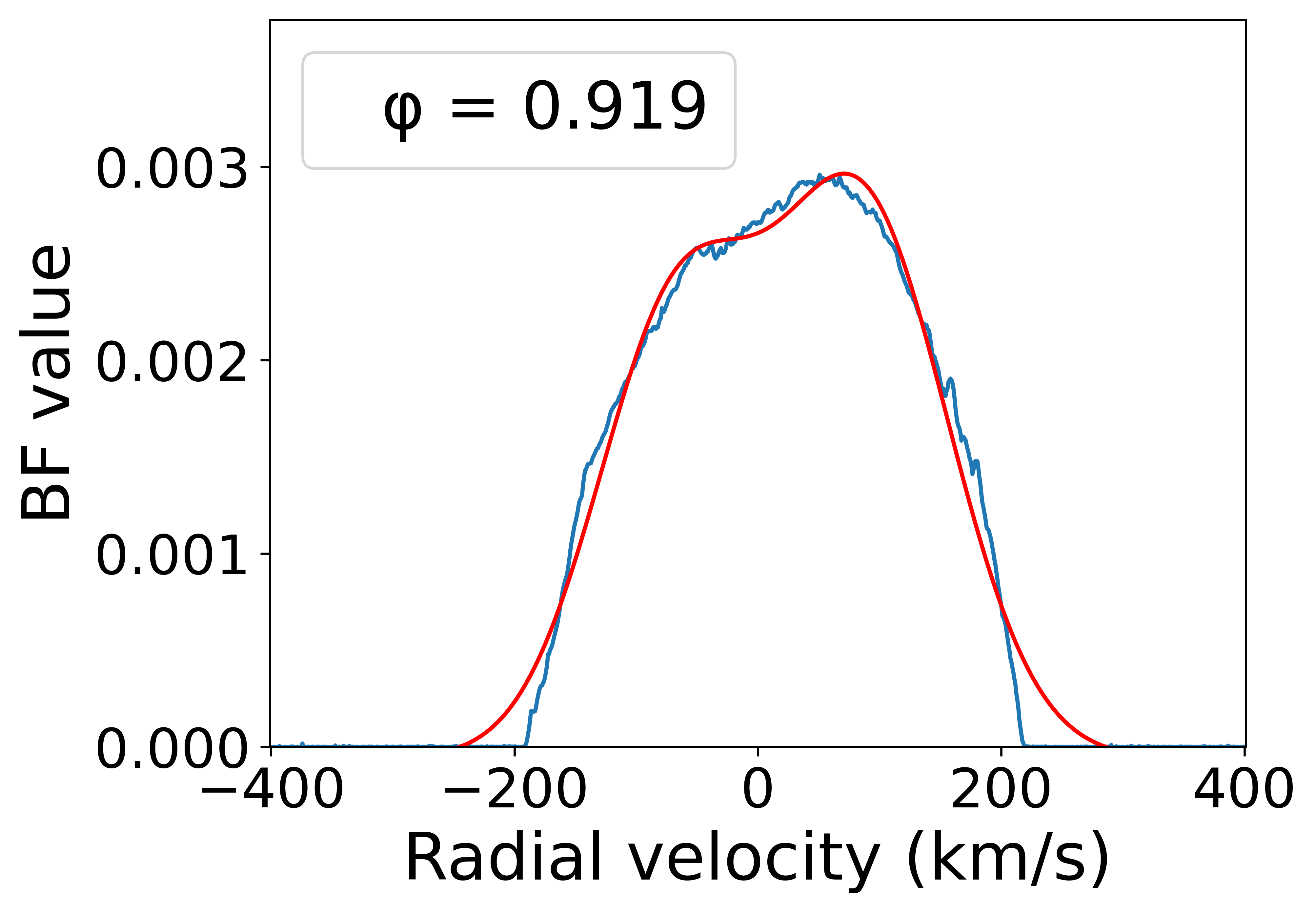} 
\caption{Representative BF profiles and corresponding Gaussian fits at selected orbital phases for the AL~Lep system.}
\label{fig:allep_bf}
\end{figure}

\begin{figure}
\centering
\includegraphics[width=0.22\textwidth]{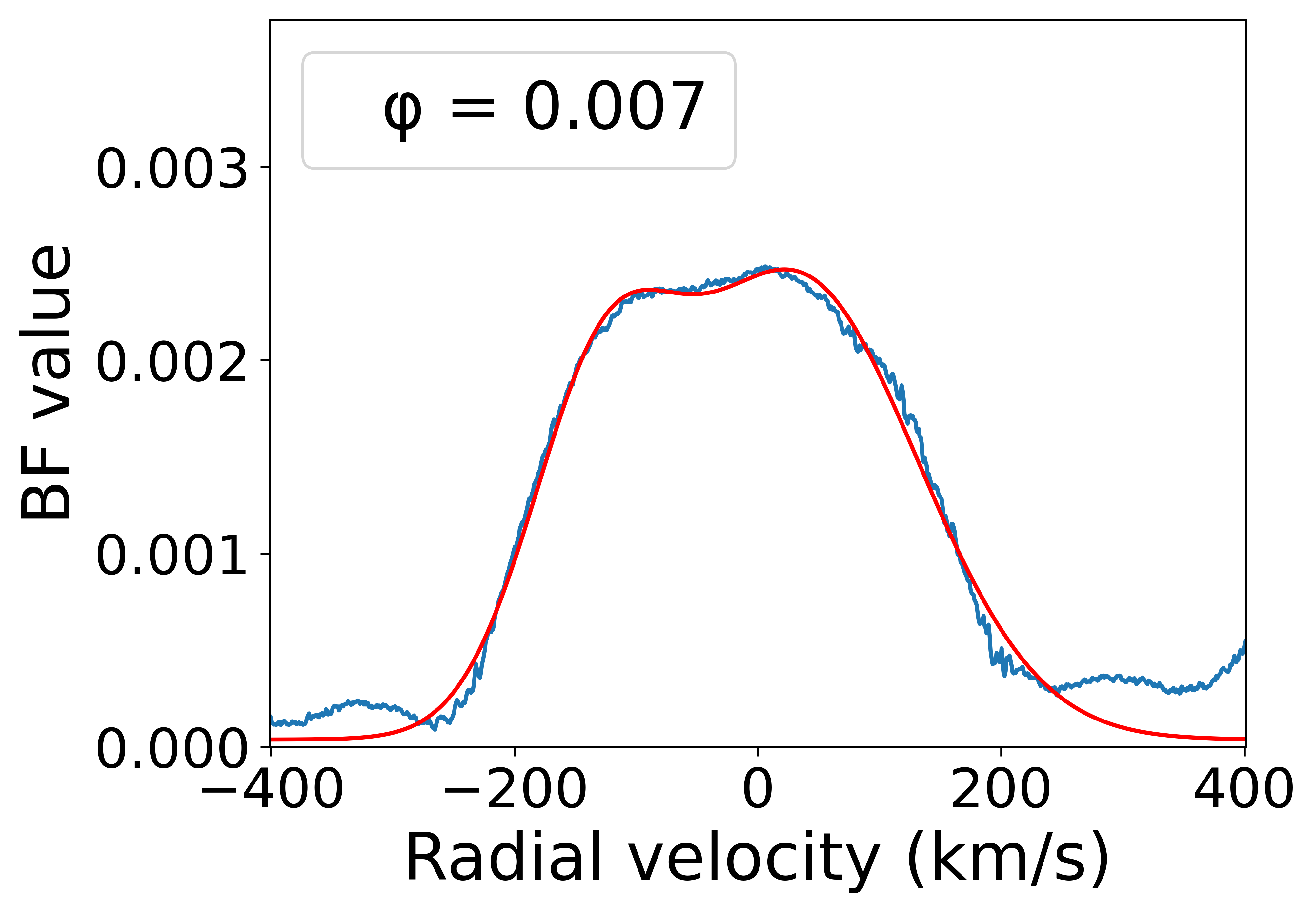}
\includegraphics[width=0.22\textwidth]{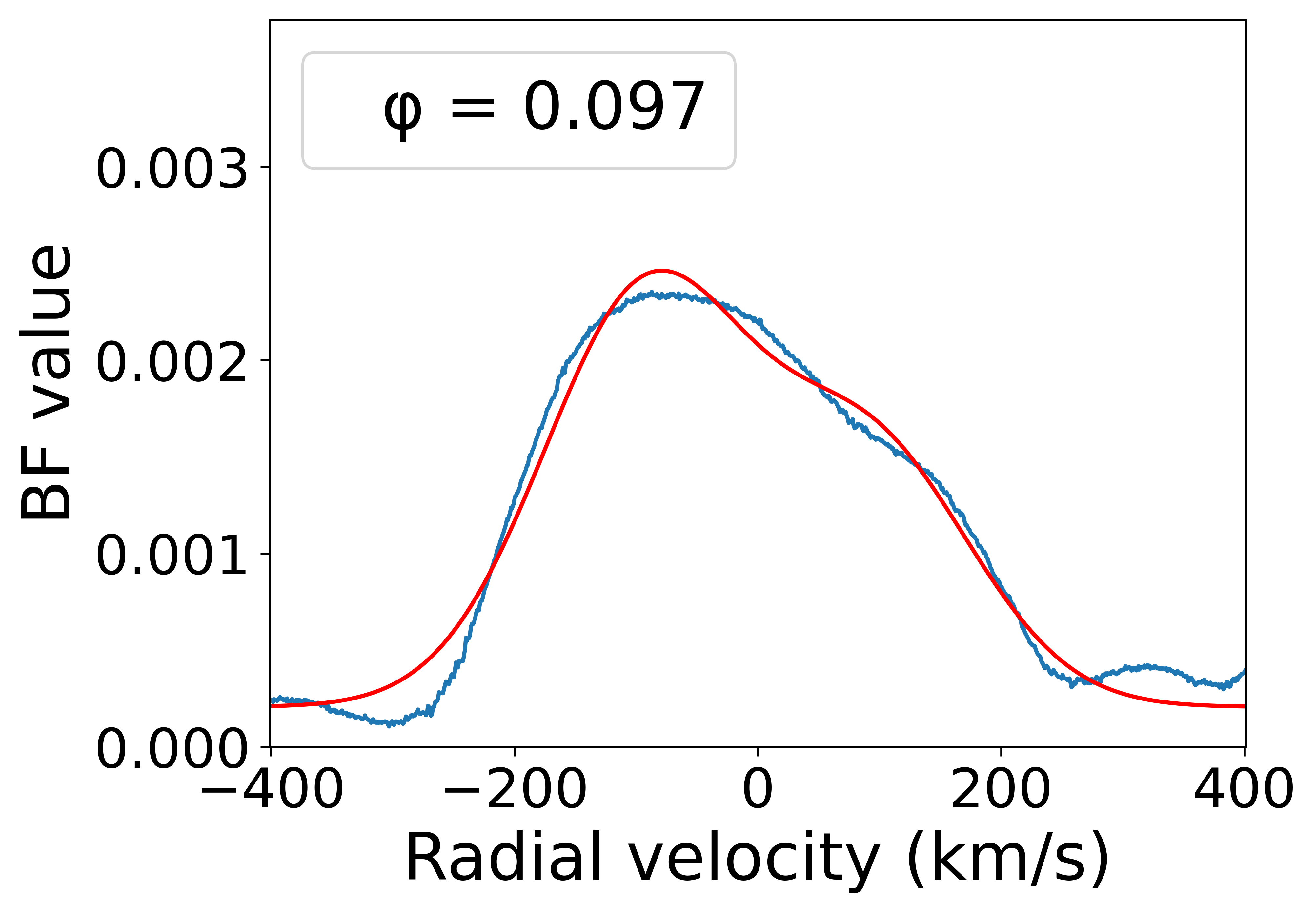}
\includegraphics[width=0.22\textwidth]{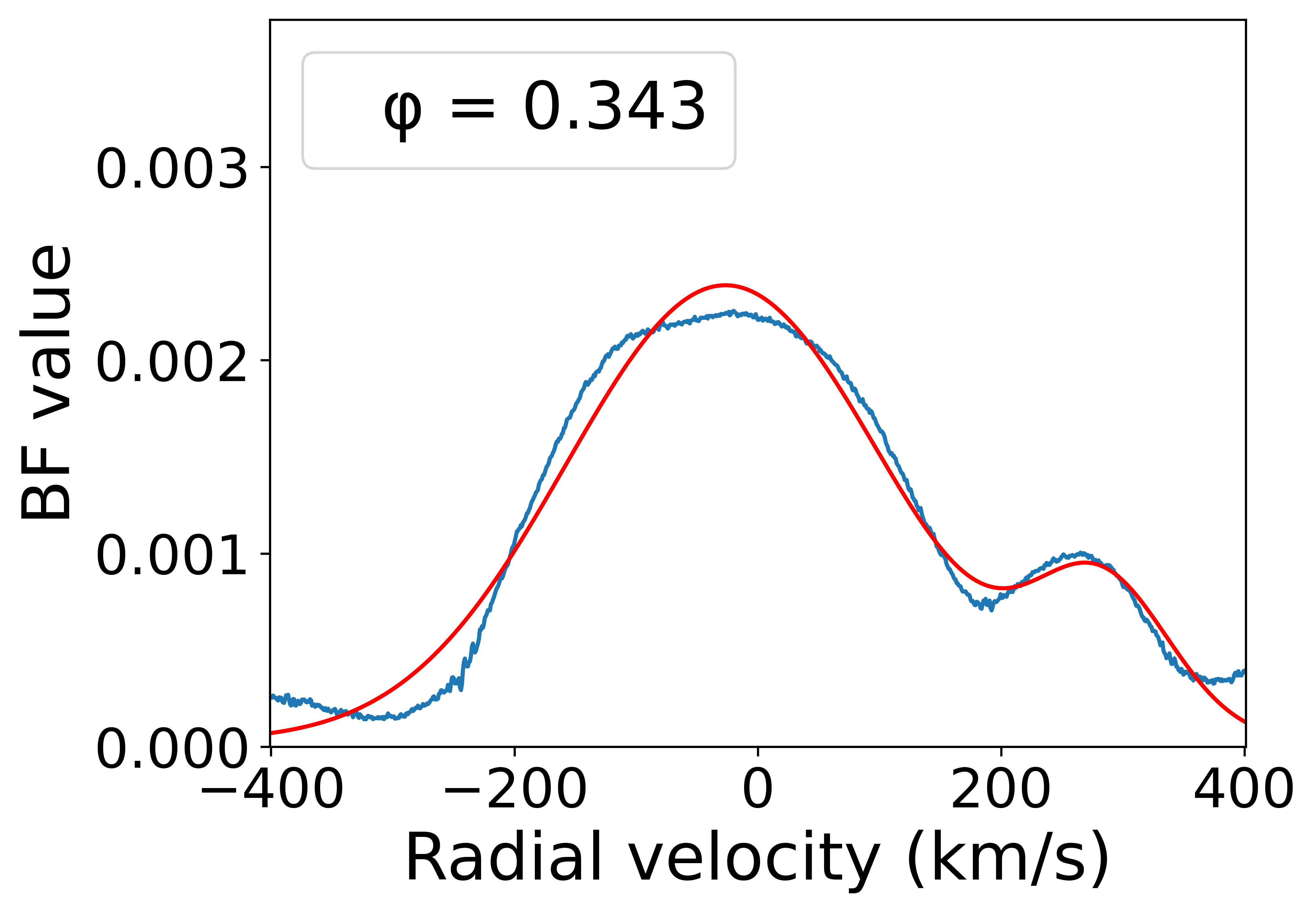}
\includegraphics[width=0.22\textwidth]{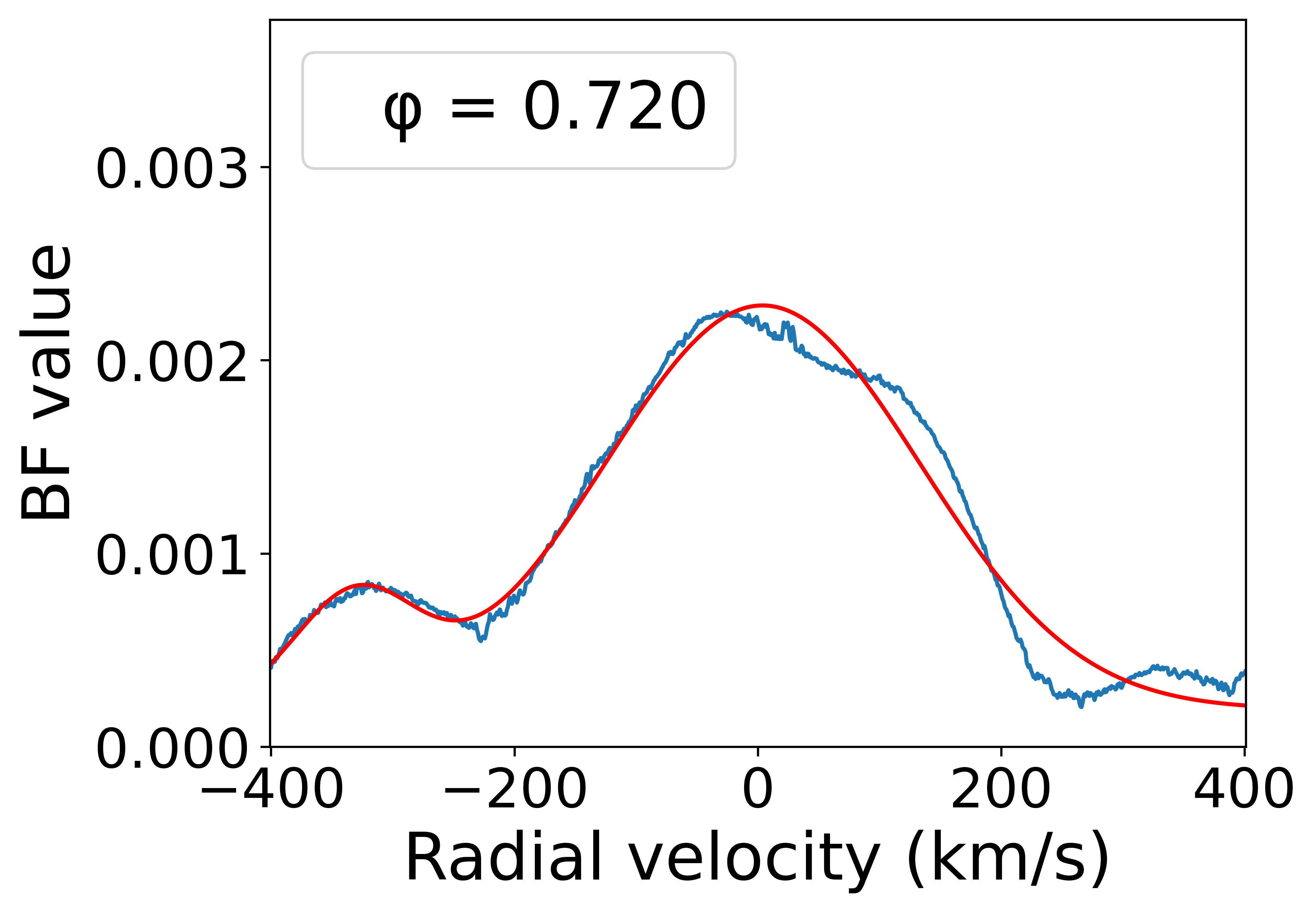}
\caption{Representative BF profiles and corresponding Gaussian fits at selected orbital phases for the ASAS J082243+1927.0 system.}
\label{fig:asasj0822_bf}
\end{figure}

\begin{figure}
\centering
\includegraphics[width=0.22\textwidth]{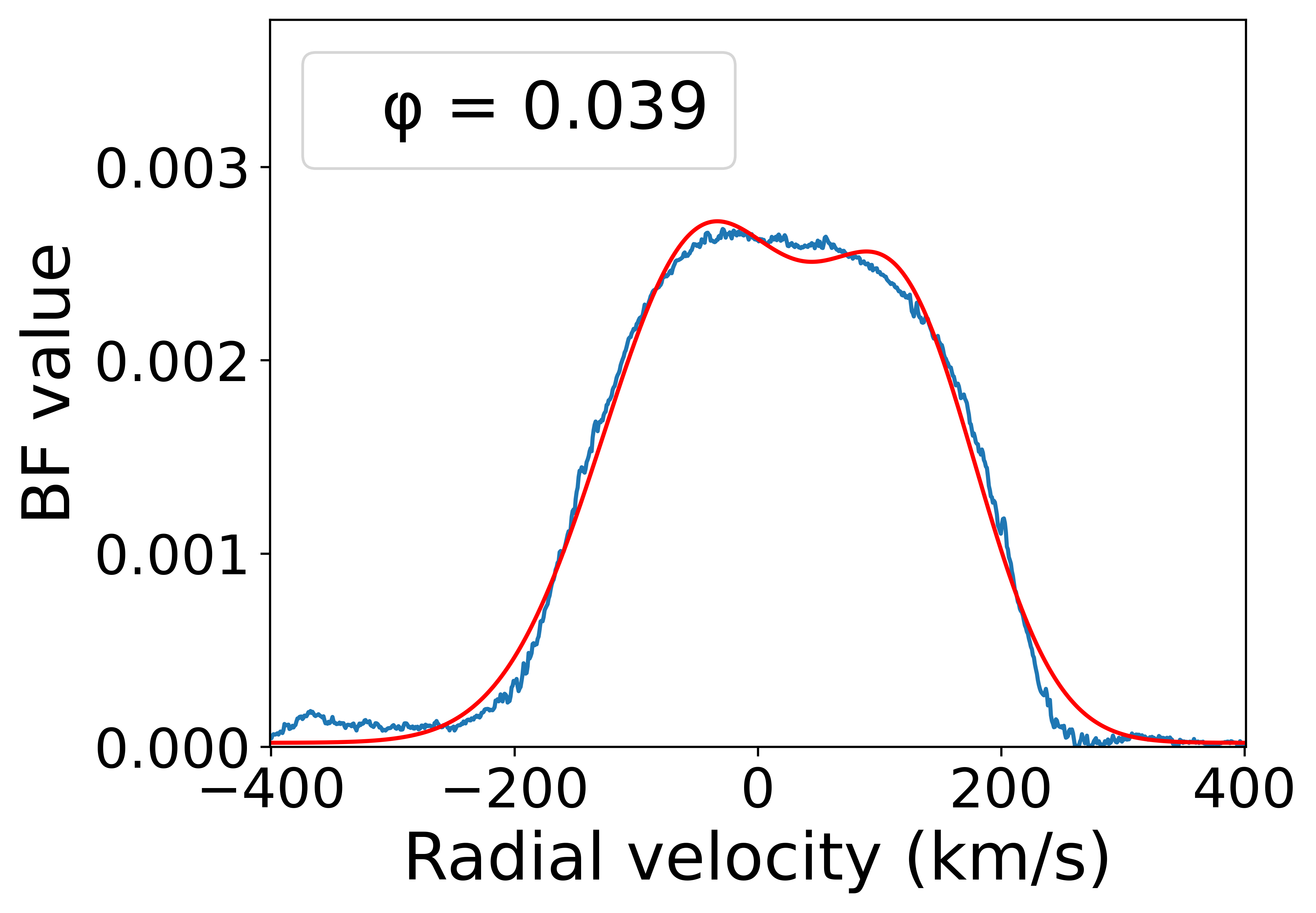}
\includegraphics[width=0.22\textwidth]{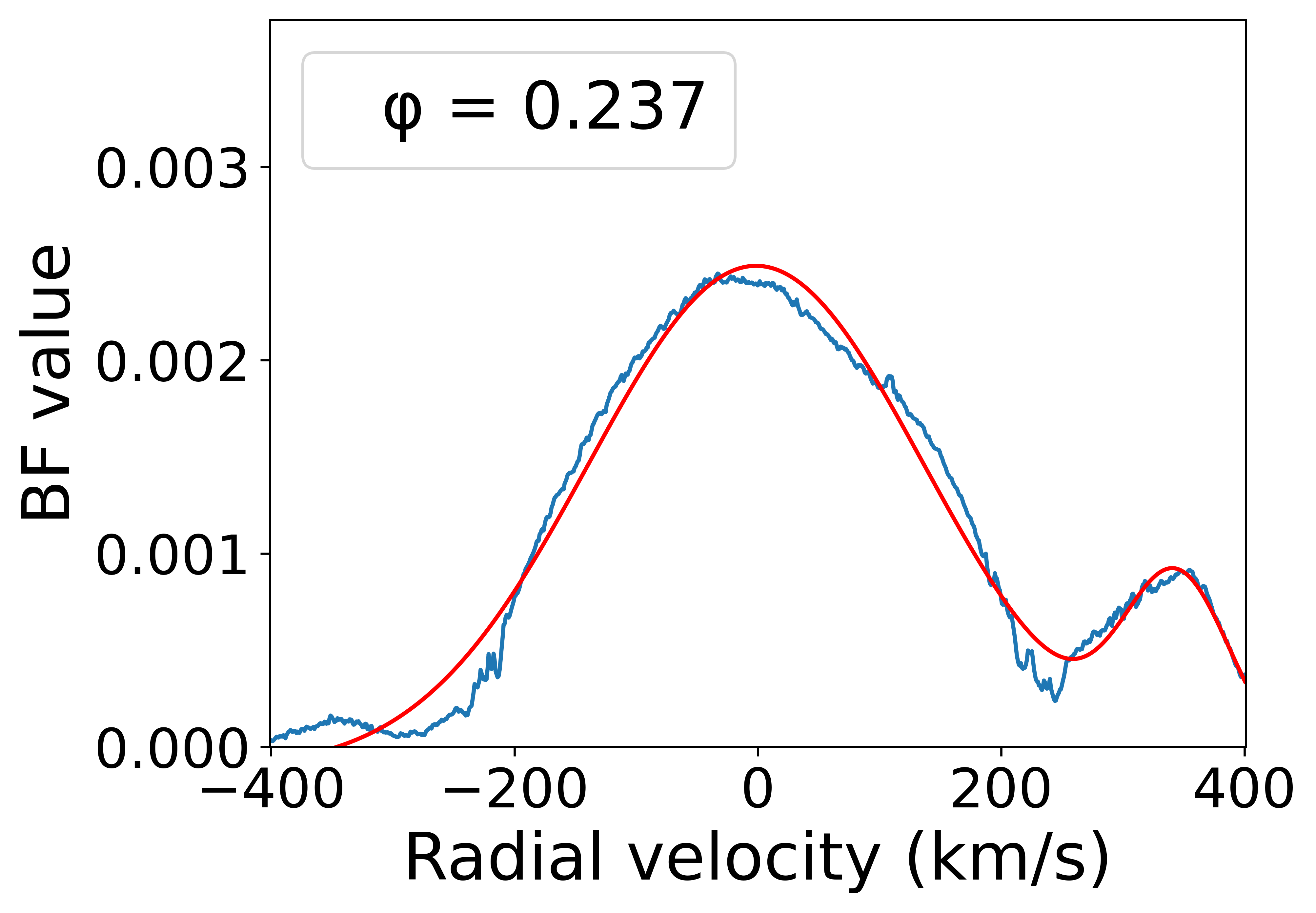}
\includegraphics[width=0.22\textwidth]{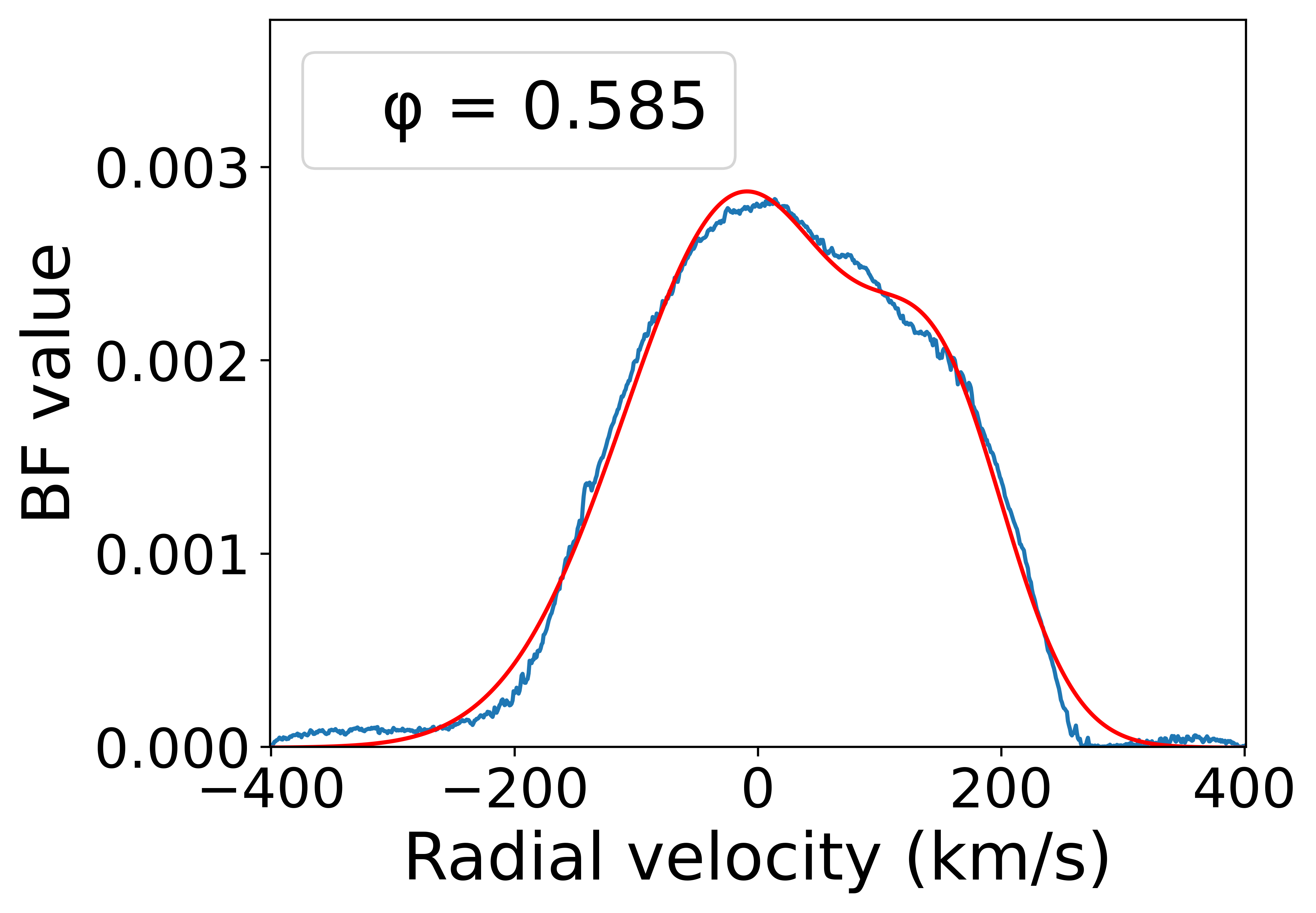}
\includegraphics[width=0.22\textwidth]{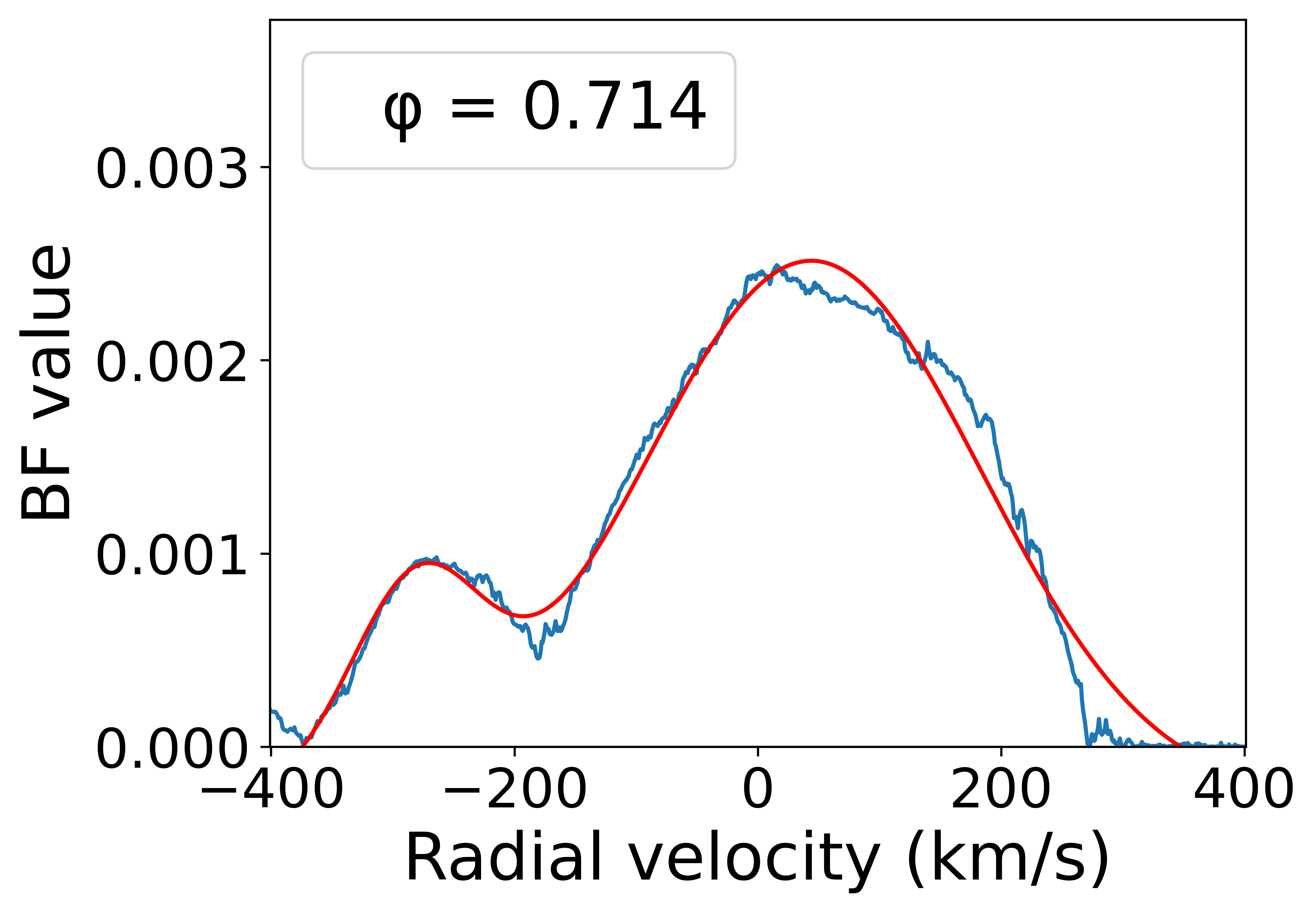}
\caption{Representative BF profiles and corresponding Gaussian fits at selected orbital phases for the ZZ~PsA system.}
\label{fig:zzpsa_bf}
\end{figure}

\begin{table}[ht!]
\caption{Corrected radial velocities of the primary ($V_1$) and secondary ($V_2$) components for the observed spectra of AL~Lep, ASAS~J082243+1927.0, and ZZ~PsA.}
\label{rv_data}
\centering
\begin{tabular}{lccccc}
\hline\hline
System & Phase & $V_1$ & $\delta V_1$ & $V_2$ & $\delta V_2$ \\
       &       & (km~s$^{-1}$) & (km~s$^{-1}$) & (km~s$^{-1}$) & (km~s$^{-1}$)\\
\hline
AL~Lep & 0.654 & 44.20 & 0.23 & -240.90 & 1.07 \\
       & 0.919 & 36.74 & 0.73 & -- & -- \\
       & 0.306 & -18.75 & 0.36 & 290.32 & 1.01 \\
       & 0.773 & 51.74 & 0.52 & -266.24 & 1.15 \\
       & 0.074 & -30.95 & 0.07 & 133.92 & 2.14 \\
       & 0.599 & 24.96 & 0.64 & -- & -- \\
       & 0.188 & -24.34 & 0.33 & 295.80 & 1.02 \\
       & 0.354 & -14.36 & 0.16 & 258.87 & 0.57 \\
       & 0.727 & 48.45 & 0.52 & -274.34 & 1.07 \\
       & 0.413 & 1.02 & 0.54 & -- & -- \\
       & 0.639 & 40.45 & 0.46 & -197.27 & 1.15 \\
       & 0.960 & 25.30 & 0.72 & -36.59 & 1.05 \\
       & 0.467 & -- & -- & -- & -- \\
       & 0.796 & 49.20 & 0.55 & -263.56 & 1.23 \\
\hline
ASAS~J082243+1927.0 & 0.017 & -17.15 & 0.93 & -- & -- \\
       & 0.530 & 6.80 & 0.24 & -75.25 & 0.80 \\
       & 0.956 & 32.75 & -- & -70.96 & 1.77 \\
       & 0.694 & 24.99 & 0.51 & -302.28 & 2.10 \\
       & 0.343 & -25.01 & 1.05 & 257.42 & 0.29 \\
       & 0.152 & -27.73 & 1.05 & 254.68 & 2.33 \\
       & 0.428 & -9.52 & 0.60 & 151.03 & 5.50 \\
       & 0.634 & 23.11 & 0.39 & -256.93 & 0.91 \\
       & 0.720 & 25.62 & 0.12 & -308.34 & 1.70 \\
       & 0.097 & -- & -- & -- & -- \\
       & 0.007 & -10.37 & 0.62 & -- & -- \\
       & 0.582 & 16.46 & 0.43 & -131.20 & 1.09 \\
       & 0.425 & -14.37 & 0.93 & 156.27 & 5.50 \\
       & 0.001 & -10.42 & 0.72 & -- & -- \\
\hline
ZZ~PsA & 0.445 & -19.65 & 0.37 & -- & -- \\
       & 0.364 & -17.04 & 0.74 & 249.04 & 0.70 \\
       & 0.812 & 20.49 & 0.78 & -305.96 & 1.77 \\
       & 0.444 & -14.91 & 0.39 & 145.05 & 1.75 \\
       & 0.379 & -17.39 & 2.00 & 258.31 & 0.83 \\
       & 0.237 & -23.26 & 0.30 & 323.80 & 0.95 \\
       & 0.133 & -28.30 & 0.56 & 222.46 & 2.62 \\
       & 0.591 & 4.36 & 0.52 & -- & -- \\
       & 0.585 & 1.76 & 0.54 & -- & -- \\
       & 0.031 & -- & -- & -- & -- \\
       & 0.714 & 17.74 & 0.66 & -305.18 & 1.26 \\
       & 0.039 & -- & -- & 99.30 & 1.25 \\
       & 0.072 & -22.52 & 1.01 & 149.71 & 3.00 \\
       & 0.718 & 23.75 & 0.74 & -309.91 & 3.00 \\
\hline
\end{tabular}
\end{table}

\begin{table}[ht!]
\caption{Final spectroscopic parameters of the three systems.}
\label{tab:spectr_param}
\centering
\begin{tabular}{lcccccc}
\hline\hline
System & $K_1$ & $K_2$ & $\gamma$ & $q_{\rm sp}$ & $T_{\rm eff}$ & $[\mathrm{Fe/H}]$ \\
& (km~s$^{-1}$) & (km~s$^{-1}$) & (km~s$^{-1}$) & & (K) & \\
\hline
AL~Lep & $-38.51 \pm 4.37$ & $292.69 \pm 4.37$ & $14.71 \pm 2.57$ & $0.132 \pm 0.015$ & $6285 \pm 133$ & $-0.056 \pm 0.030$ \\
ASAS~J082243+1927.0 & $-30.17 \pm 5.76$ & $318.34 \pm 5.76$ & $1.89 \pm 2.75$ & $0.095 \pm 0.018$ & $5880 \pm 132$ & $-0.524 \pm 0.038$ \\
ZZ~PsA & $-25.78 \pm 5.95$ & $329.29 \pm 5.95$ & $1.24 \pm 3.34$  
& $0.078 \pm 0.018$ & $5898 \pm 110$ & $-0.042 \pm 0.029$ \\
\hline
\end{tabular}
\end{table}

\begin{figure}[ht!]
\centering
\includegraphics[width=\textwidth]{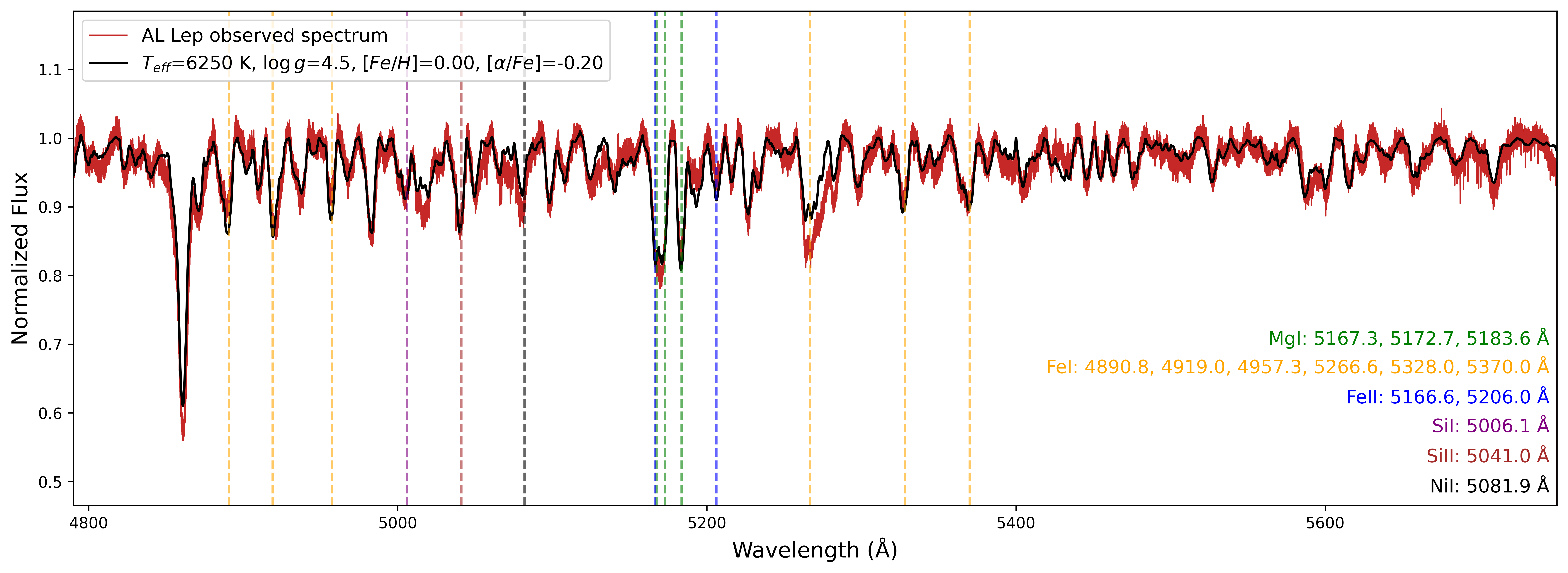}\\
\includegraphics[width=\textwidth]{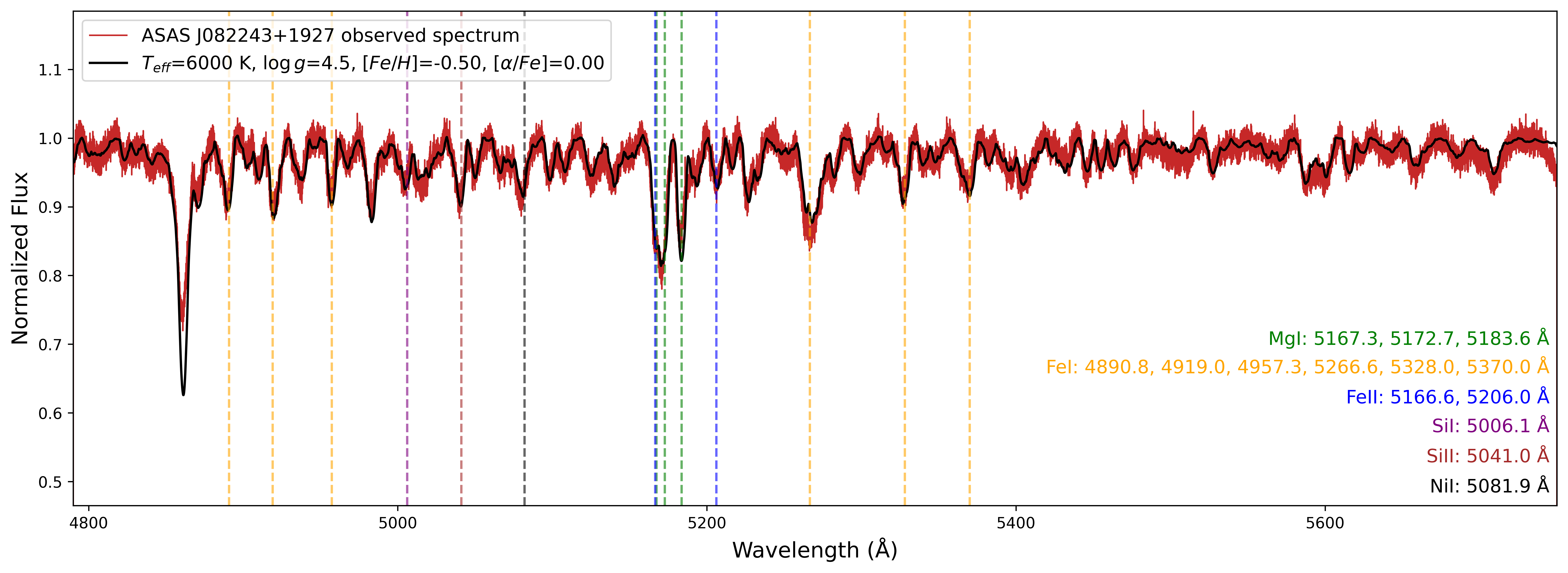} \\
\includegraphics[width=\textwidth]{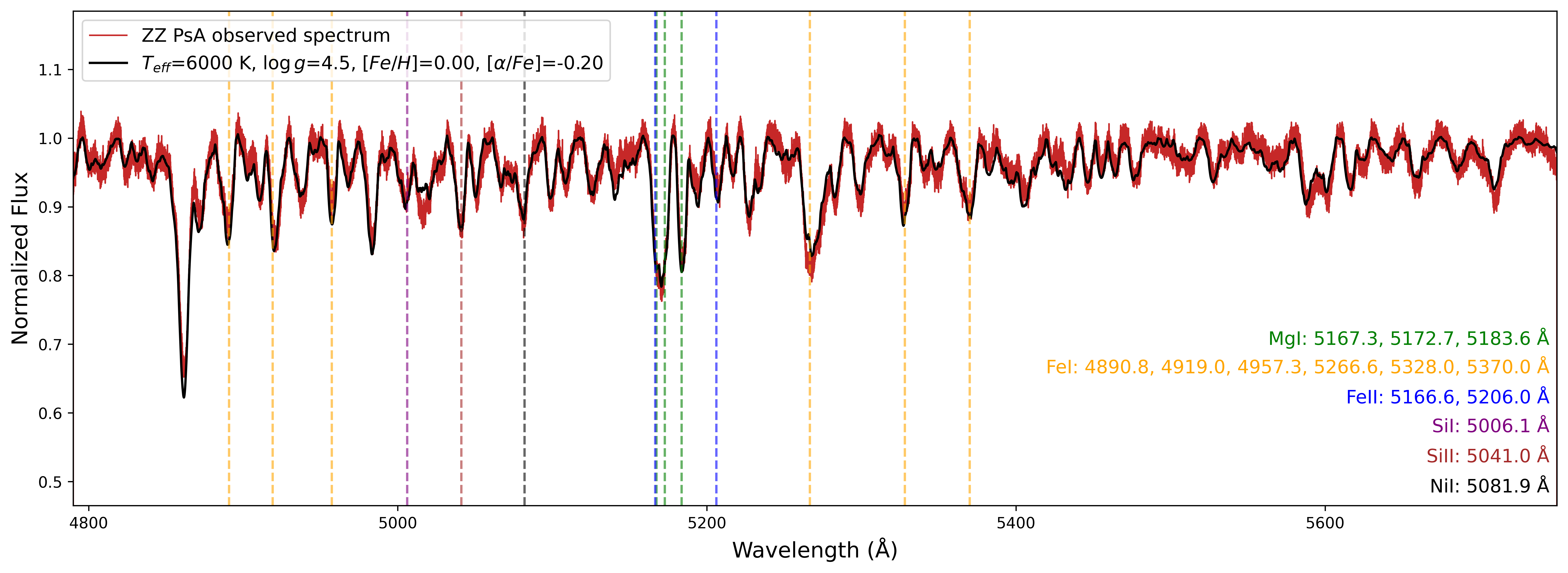} \\
\caption{Observed spectrum of the three systems (red) overlaid with the best-fitting synthetic spectrum (black) for the adopted atmospheric parameters. Vertical dashed lines mark the selected absorption lines used in the analysis, corresponding to Fe\,I, Fe\,II, Mg\,I, Si\,I, Si\,II, and Ni\,I transitions.}
\label{fig:lines}
\end{figure}

\twocolumn

\begin{figure}[ht!]
\centering
\includegraphics[width=0.49\textwidth]{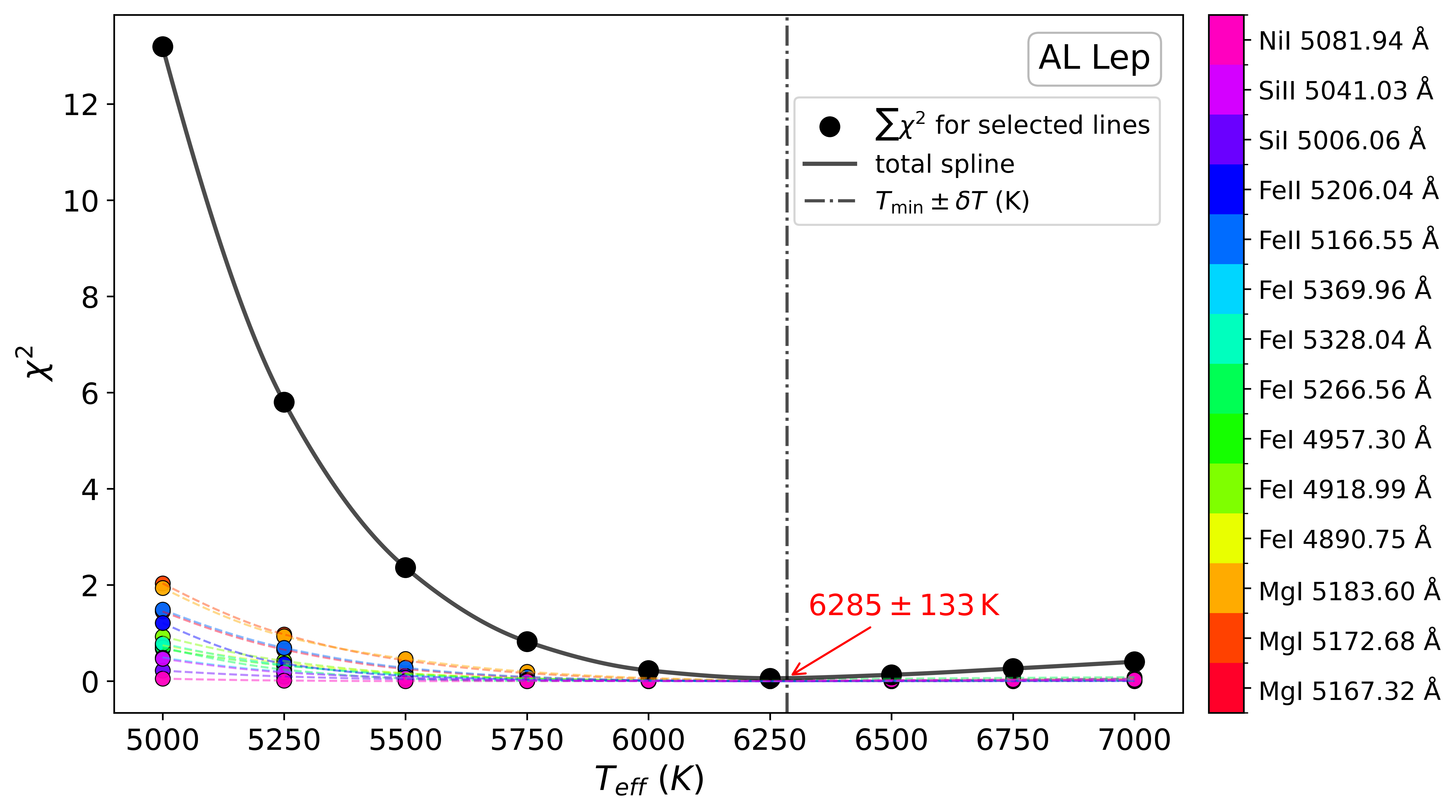}\\
\includegraphics[width=0.49\textwidth]{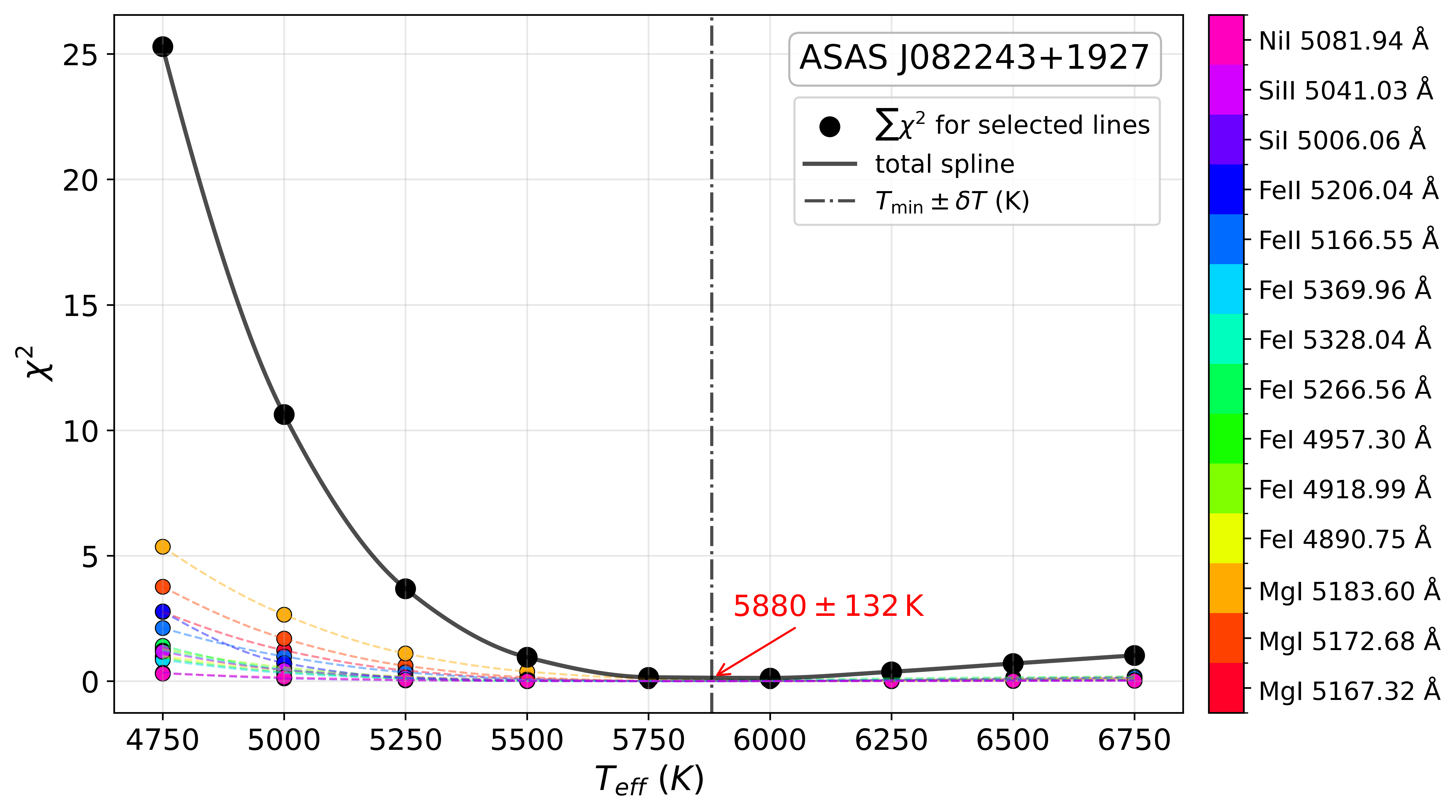} \\
\includegraphics[width=0.49\textwidth]{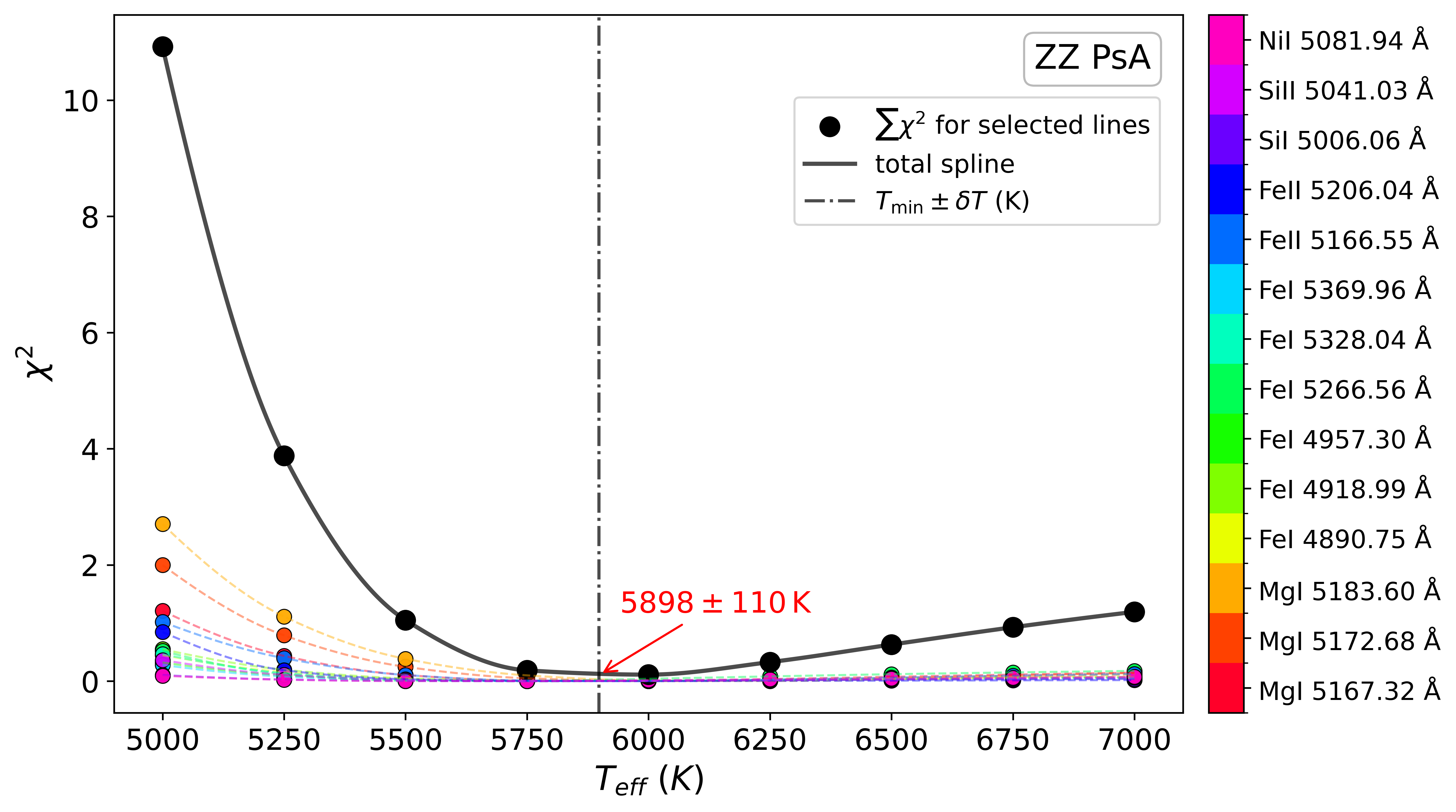} \\
\caption{Temperature dependence of the mismatch metric for the three targets. Coloured points and dashed curves show the $\chi^2$ values for the individual selected spectral lines. Black points and solid curve represent the summed mismatch, $\sum \chi^2$. The vertical dash-dotted line marks the temperature corresponding to the minimum of the total mismatch, and the adopted effective temperature.}
\label{fig:chi2_teff}
\end{figure}

\begin{figure}[ht!]
\centering
\includegraphics[width=0.49\textwidth]{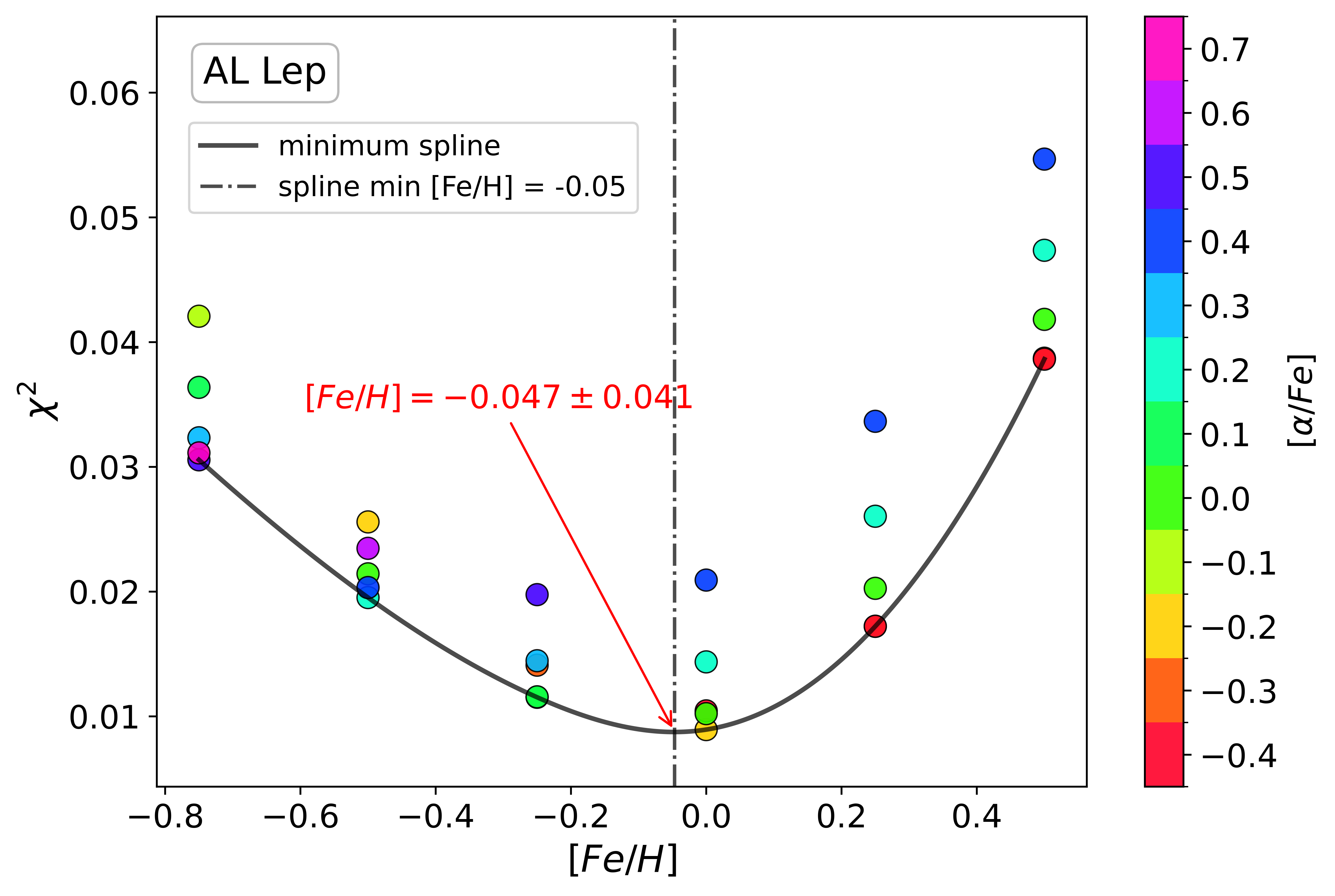} \\
\includegraphics[width=0.49\textwidth]{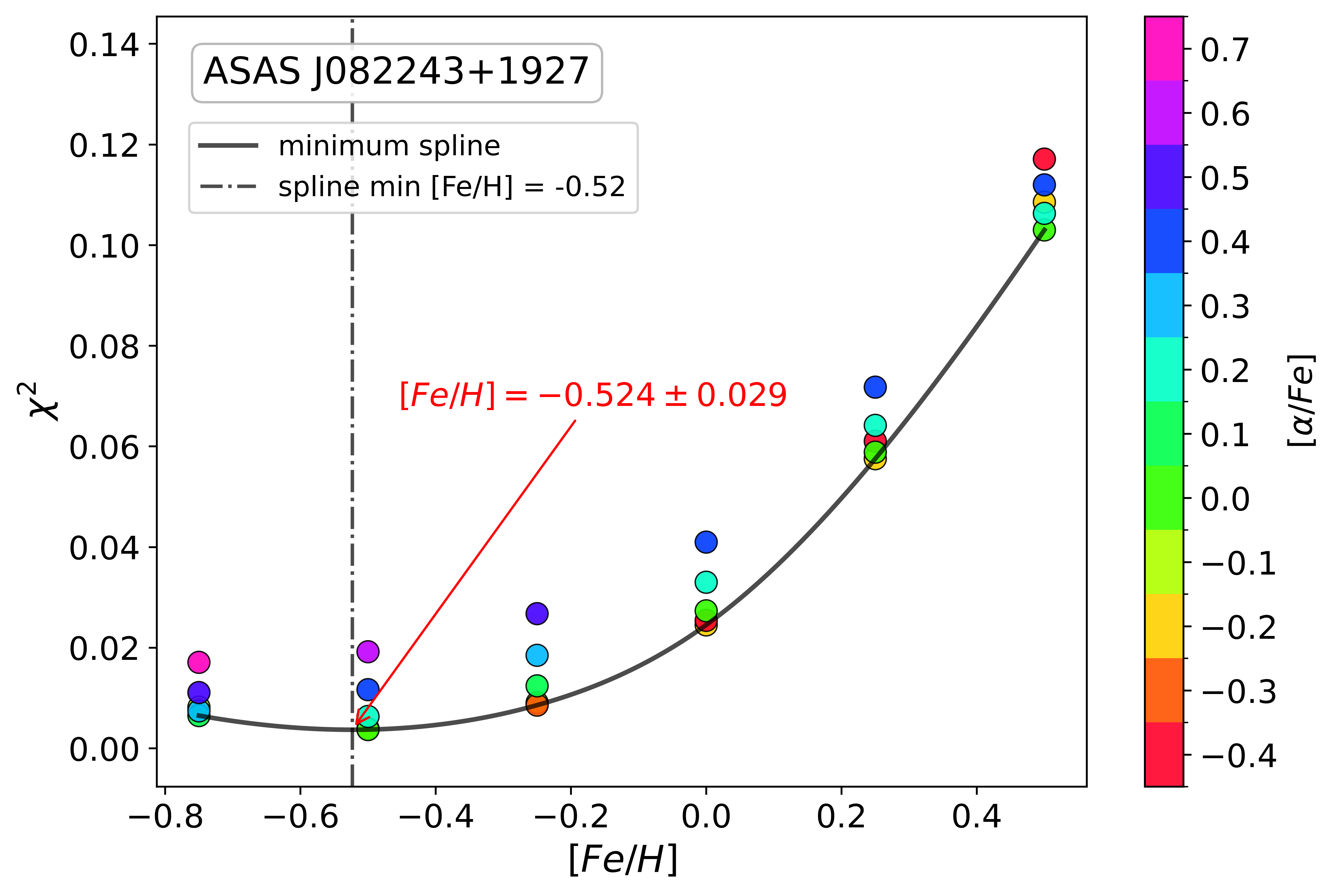} \\
\includegraphics[width=0.49\textwidth]{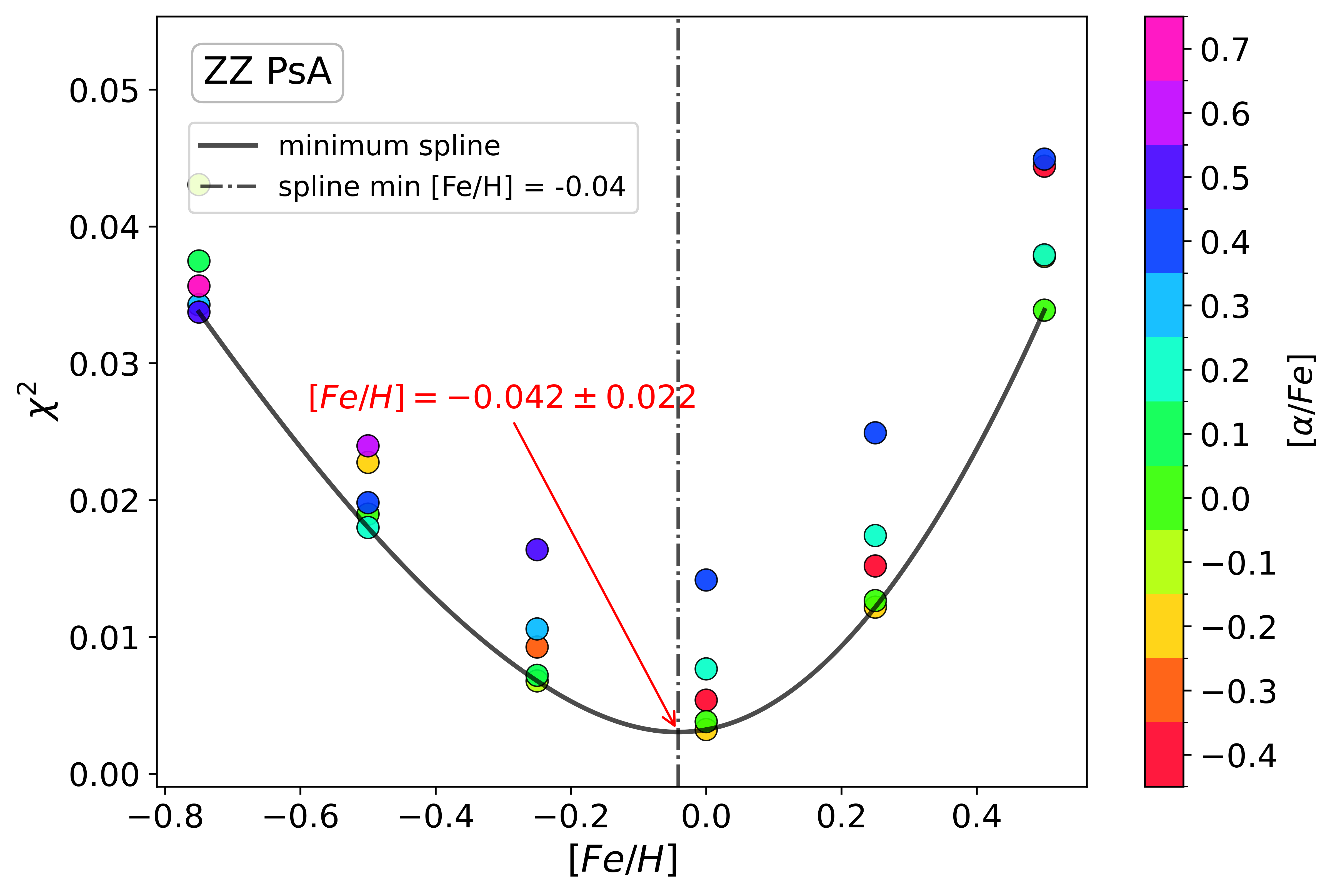} \\
\caption{Metallicity dependence of the mismatch metric for the three systems. Coloured points correspond to the synthetic spectra in the metallicity grid, with colour indicating $[\alpha/Fe]$ values. Black curve shows the spline interpolation through the minimum $\chi^2$ value at each $[Fe/H]$, after selecting the best available $[\alpha/Fe]$ model. Vertical dash-dotted line marks the minimum of the interpolated curve and gives the adopted spectroscopic metallicity.}
\label{fig:metal}
\end{figure}

\onecolumn
\FloatBarrier

\section{Light curve modelling parameters}

\begin{table}[ht!]
\caption{Final LC model parameters for AL~Lep, ASAS~J082243+1927.0, and ZZ~PsA.}
\label{tab:model_parameters}
\centering
\setlength{\tabcolsep}{2.5pt}
\begin{tabular}{lcccccc}
\hline\hline
 & \multicolumn{2}{c}{AL~Lep} & \multicolumn{2}{c}{ASAS~J082243+1927.0} & \multicolumn{2}{c}{ZZ~PsA} \\
\hline
Parameter & B,V,R,I,C & S & B,V,R,I,C & S & B,V,R,I,C & S \\
\hline
$f$ (\%) & \multicolumn{2}{c}{22 $\pm$ 2} & \multicolumn{2}{c}{25 $\pm$ 5} & \multicolumn{2}{c}{98 $\pm$ 2} \\
$i$ ($^\circ$) & \multicolumn{2}{c}{75.0 $\pm$ 0.2} & \multicolumn{2}{c}{73.4 $\pm$ 0.2} & \multicolumn{2}{c}{73.8 $\pm$ 0.3} \\
$T_1$ (K) & \multicolumn{2}{c}{6285*} & \multicolumn{2}{c}{5880*} & \multicolumn{2}{c}{5898*} \\
$T_2$ (K) & \multicolumn{2}{c}{6141 $\pm$ 8} & \multicolumn{2}{c}{6006 $\pm$ 12} & \multicolumn{2}{c}{6083 $\pm$ 10} \\
$\Omega_1 = \Omega_2$ & \multicolumn{2}{c}{2.038 $\pm$ 0.002} & \multicolumn{2}{c}{1.954 $\pm$ 0.004} & \multicolumn{2}{c}{1.840 $\pm$ 0.001} \\
$q_{\rm ph}$ & \multicolumn{2}{c}{0.1332 $\pm$ 0.0005} & \multicolumn{2}{c}{0.1041 $\pm$ 0.0011} & \multicolumn{2}{c}{0.0789 $\pm$ 0.0003} \\
$L_1/(L_1+L_2)_B$ & 0.8669 $\pm$ 0.0006 & -- & 0.8684 $\pm$ 0.0006 & -- & 0.8739 $\pm$ 0.0004 & -- \\
$L_1/(L_1+L_2)_V$ & 0.8658 $\pm$ 0.0006 & -- & 0.8701 $\pm$ 0.0005 & -- & 0.8762 $\pm$ 0.0003 & -- \\
$L_1/(L_1+L_2)_R$ & 0.8646 $\pm$ 0.0005 & -- & 0.8715 $\pm$ 0.0004 & -- & 0.8783 $\pm$ 0.0003 & -- \\
$L_1/(L_1+L_2)_I$ & 0.8628 $\pm$ 0.0004 & -- & 0.8734 $\pm$ 0.0001 & -- & 0.8777 $\pm$ 0.0001 & -- \\
$L_1/(L_1+L_2)_C$ & 0.8663 $\pm$ 0.0001 & -- & 0.8709 $\pm$ 0.0001 & -- & 0.8966 $\pm$ 0.0003 & -- \\
$L_1/(L_1+L_2)_S$ & -- & 0.8552 $\pm$ 0.0001 & -- & 0.8786 $\pm$ 0.0005 & -- & 0.886 $\pm$ 0.001 \\
$L_2/(L_1+L_2)_B$ & 0.1331 $\pm$ 0.0006 & -- & 0.1316 $\pm$ 0.0006 & -- & 0.1261 $\pm$ 0.0004 & -- \\
$L_2/(L_1+L_2)_V$ & 0.1342 $\pm$ 0.0006 & -- & 0.1299 $\pm$ 0.0005 & -- & 0.1238 $\pm$ 0.0003 & -- \\
$L_2/(L_1+L_2)_R$ & 0.1354 $\pm$ 0.0005 & -- & 0.1285 $\pm$ 0.0005 & -- & 0.1217 $\pm$ 0.0003 & -- \\
$L_2/(L_1+L_2)_I$ & 0.1372 $\pm$ 0.0004 & -- & 0.1266 $\pm$ 0.0004 & -- & 0.1189 $\pm$ 0.0002 & -- \\
$L_2/(L_1+L_2)_C$ & 0.1337 $\pm$ 0.0001 & -- & 0.1291 $\pm$ 0.0001 & -- & 0.1034 $\pm$ 0.0003 & -- \\
$L_2/(L_1+L_2)_S$ & -- & 0.145 $\pm$ 0.001 & -- & 0.1214 $\pm$ 0.0005 & -- & 0.1223 $\pm$ 0.0001 \\
$r_{\rm 1,volume}$ & \multicolumn{2}{c}{0.5648 $\pm$ 0.0008} & \multicolumn{2}{c}{0.5844 $\pm$ 0.0018} & \multicolumn{2}{c}{0.6227 $\pm$ 0.0004} \\
$r_{\rm 2,volume}$ & \multicolumn{2}{c}{0.2323 $\pm$ 0.0009} & \multicolumn{2}{c}{0.2169 $\pm$ 0.0019} & \multicolumn{2}{c}{0.2235 $\pm$ 0.0006} \\
\hline
\multicolumn{7}{c}{\it Spot parameters} \\
\hline
Latitude ($^\circ$) & 139.34 $\pm$ 6.69 & 156.72 $\pm$ 0.03 & 137.48 $\pm$ 3.81 & 162.91 $\pm$ 0.26 & 95.80 $\pm$ 5.03 & 129.64 $\pm$ 0.06 \\
Longitude ($^\circ$) & 356.46 $\pm$ 5.17 & 12.29 $\pm$ 0.08 & 347.43 $\pm$ 1.45 & 200.17 $\pm$ 0.83 & 278.56 $\pm$ 0.98 & 297.14 $\pm$ 0.14 \\
Radius ($^\circ$) & 25.14 $\pm$ 3.85 & 25.04 $\pm$ 0.03 & 37.50 $\pm$ 1.56 & 39.90 $\pm$ 0.06 & 19.79 $\pm$ 2.30 & 23.72 $\pm$ 0.06 \\
Temp. factor & 0.90 $\pm$ 0.05 & 0.70 $\pm$ 0.01 & 0.87 $\pm$ 0.02 & 0.71 $\pm$ 0.01 & 0.85 $\pm$ 0.02 & 0.70 $\pm$ 0.01 \\
\hline
\end{tabular}
\tablefoot{Asterisks indicate fixed primary temperature values. Fractional luminosity and radius values are dimensionless.}
\end{table}

\FloatBarrier
\twocolumn
\end{appendix}

\end{document}